\documentclass[twocolumn]{aastex63}
\usepackage{CJK}

\newcommand\N[1]{{\color{teal} \bf #1}}
\usepackage{soul}

\shorttitle{Fast-evolving Type II SN 2023ufx}
\shortauthors{Ravi et al.}
\graphicspath{{./}{figures/}}

\usepackage{tablefootnote}
\usepackage{afterpage}
\usepackage{longtable}
\usepackage{xcolor}
\usepackage{graphics}
\usepackage{epsf}
\usepackage{enumitem}
\usepackage{threeparttable} 
\usepackage{amsmath}
\usepackage{appendix}
\begin{document}

\title{Luminous Type II Short-Plateau SN 2023ufx: Asymmetric Explosion of a Partially-Stripped Massive Progenitor}

\correspondingauthor{Aravind Pazhayath Ravi}
\email{apazhayathravi@ucdavis.edu}

\author[0000-0002-7352-7845]{Aravind P. Ravi}
\affiliation{Department of Physics and Astronomy, University of California, 1 Shields Avenue, Davis, CA 95616-5270, USA} 

\author[0000-0001-8818-0795]{Stefano Valenti}
\affiliation{Department of Physics and Astronomy, University of California, 1 Shields Avenue, Davis, CA 95616-5270, USA}

\author[0000-0002-7937-6371]{Yize Dong \begin{CJK*}{UTF8}{gbsn}(董一泽)\end{CJK*}}
\affiliation{Department of Physics and Astronomy, University of California, 1 Shields Avenue, Davis, CA 95616-5270, USA}
\affiliation{Center for Astrophysics \textbar{} Harvard \& Smithsonian, 60 Garden Street, Cambridge, MA 02138-1516, USA}

\author[0000-0002-1125-9187]{Daichi Hiramatsu}
\affiliation{Center for Astrophysics \textbar{} Harvard \& Smithsonian, 60 Garden Street, Cambridge, MA 02138-1516, USA}
\affiliation{The NSF AI Institute for Artificial Intelligence and Fundamental Interactions, USA}

\author[0000-0003-4800-2737]{Stan Barmentloo}
\affiliation{The Oskar Klein Centre, Department of Astronomy, Stockholm University, AlbaNova, SE-10691 Stockholm, Sweden}

\author[0000-0001-8005-4030]{Anders Jerkstrand}
\affiliation{The Oskar Klein Centre, Department of Astronomy, Stockholm University, AlbaNova, SE-10691 Stockholm, Sweden}

\author[0000-0002-4924-444X]{K.\ Azalee Bostroem}
\altaffiliation{LSST-DA Catalyst Fellow}
\affiliation{Steward Observatory, University of Arizona, 933 North Cherry Avenue, Tucson, AZ 85721-0065, USA}

\author[0000-0002-0744-0047]{Jeniveve Pearson}
\affiliation{Steward Observatory, University of Arizona, 933 North Cherry Avenue, Tucson, AZ 85721-0065, USA}

\author[0000-0002-4022-1874]{Manisha Shrestha}
\affil{Steward Observatory, University of Arizona, 933 North Cherry Avenue, Tucson, AZ 85721-0065, USA}

\author[0000-0003-0123-0062]{Jennifer E.\ Andrews}
\affiliation{Gemini Observatory, 670 North A`ohoku Place, Hilo, HI 96720-2700, USA}

\author[0000-0003-4102-380X]{David J.\ Sand}
\affiliation{Steward Observatory, University of Arizona, 933 North Cherry Avenue, Tucson, AZ 85721-0065, USA}

\author[0000-0002-0832-2974]{Griffin Hosseinzadeh} \affiliation{Department of Astronomy \& Astrophysics, University of California, San Diego, 9500 Gilman Drive, MC 0424, La Jolla, CA 92093-0424, USA}

\author[0000-0001-9589-3793]{Michael Lundquist}
\affiliation{W.~M.~Keck Observatory, 65-1120 M\=amalahoa Highway, Kamuela, HI 96743-8431, USA}

\author[0000-0003-2744-4755]{Emily Hoang}
\affil{Department of Physics and Astronomy, University of California, 1 Shields Avenue, Davis, CA 95616-5270, USA}

\author[0009-0008-9693-4348]{Darshana Mehta}
\affiliation{Department of Physics and Astronomy, University of California, 1 Shields Avenue, Davis, CA 95616-5270, USA}

\author[0000-0002-7015-3446]{Nicol\'as Meza Retamal}
\affiliation{Department of Physics and Astronomy, University of California, 1 Shields Avenue, Davis, CA 95616-5270, USA}

\author[0009-0001-3106-0917]{Aidan Martas}\affiliation{Department of Astronomy, University of California, Berkeley, CA 94720-3411, USA}\affiliation{Department of Physics and Astronomy, University of California, 1 Shields Avenue, Davis, CA 95616-5270, USA}

\author[0000-0001-8738-6011]{Saurabh W.\ Jha}\affiliation{Department of Physics and Astronomy, Rutgers, the State University of New Jersey,\\136 Frelinghuysen Road, Piscataway, NJ 08854-8019, USA}

\author[0000-0003-0549-3281]{Daryl Janzen}
\affiliation{Department of Physics \& Engineering Physics, University of Saskatchewan, 116 Science Place, Saskatoon, SK S7N 5E2, Canada}

\author[0000-0001-8073-8731]{Bhagya Subrayan}\affiliation{Steward Observatory, University of Arizona, 933 North Cherry Avenue, Tucson, AZ 85721-0065, USA}


\newcommand{\LCO}{\affiliation{Las Cumbres Observatory, 6740 Cortona Drive, Suite 102, Goleta, CA 93117-5575, USA}}
\newcommand{\UCSB}{\affiliation{Department of Physics, University of California, Santa Barbara, CA 93106-9530, USA}}

\author[0000-0003-4253-656X]{D.\ Andrew Howell}
\LCO\UCSB

\author[0000-0001-5807-7893]{Curtis McCully}
\LCO

\author[0000-0003-4914-5625]{Joseph Farah}
\LCO 
\UCSB

\author[0000-0001-9570-0584]{Megan Newsome}
\LCO 
\UCSB

\author[0000-0003-0209-9246]{Estefania Padilla Gonzalez}
\LCO
\UCSB

\author[0000-0003-0794-5982]{Giacomo Terreran}
\LCO 
\UCSB

\author[0000-0002-1895-6639]{Moira Andrews}
\LCO
\UCSB

\author[0000-0003-3460-0103]{Alexei V. Filippenko}\affiliation{Department of Astronomy, University of California, Berkeley, CA 94720-3411, USA}

\author[0000-0001-5955-2502
]{Thomas G. Brink}\affiliation{Department of Astronomy, University of California, Berkeley, CA 94720-3411, USA}

\author[0000-0002-2636-6508]{Weikang Zheng}\affiliation{Department of Astronomy, University of California, Berkeley, CA 94720-3411, USA}

\author[0000-0002-6535-8500]{Yi Yang}\affiliation{Department of Astronomy, University of California, Berkeley, CA 94720-3411, USA}

\author[0000-0001-8764-7832]{Jozsef Vink\'o}\affiliation{CSFK Konkoly Observatory, Konkoly-Thege M. ut 15-17, Budapest, 1121, Hungary}\affiliation{Department of Optics and Quantum Electronics, University of Szeged, D\'om t\'er 9, Szeged, 6720 Hungary}\affiliation{ELTE E\"otv\"os Lor\'and University, Institute of Physics, P\'azmany P\'eter s\'et\'any 1/A, Budapest, 1117, Hungary}\affiliation{University of Texas at Austin, 1 University Station C1400, Austin, TX 78712-0259, USA}

\author[0000-0003-1349-6538]{J. Craig Wheeler}\affiliation{University of Texas at Austin, 1 University Station C1400, Austin, TX 78712-0259, USA}

\author[0000-0001-5510-2424]{Nathan Smith}\affiliation{Steward Observatory, University of Arizona, 933 North Cherry Avenue, Tucson, AZ 85721-0065, USA}

\author[0000-0003-3643-839X]{Jeonghee Rho}\affil{SETI Institute, 189 Bernardo Ave., Ste. 200, Mountain View, CA 94043, USA; jrho@seti.org}

\author[0000-0002-8770-6764]{R\'eka K\"onyves-T\'oth}\affil{CSFK Konkoly Observatory, Konkoly-Thege M. ut 15-17, Budapest, 1121, Hungary}

\author[0000-0003-2375-2064]{Claudia P. Guti\'{e}rrez}\affiliation{Institut d'Estudis Espacials de Catalunya (IEEC), Edifici RDIT, Campus UPC, 08860 Castelldefels (Barcelona), Spain} \affiliation{Institute of Space Sciences (ICE, CSIC), Campus UAB, Carrer
de Can Magrans, s/n, E-08193 Barcelona, Spain}











\begin{abstract}
We present supernova (SN)\,2023ufx, a unique Type IIP  SN with the shortest known plateau duration ($t_\mathrm{PT}$ $\sim$47 days), a luminous V-band peak ($M_{V}$ = $-$18.42 $\pm$ 0.08 mag), and a rapid early decline rate ($s1$ = 3.47 $\pm$ 0.09 mag (50 days)$^{-1}$). By comparing observed photometry to a hydrodynamic MESA+STELLA model grid, we constrain the progenitor to be a massive red supergiant with M$_\mathrm{ZAMS}$ $\simeq$19 -- 25 M$_{\odot}$. Independent comparisons with nebular spectral models also suggest an initial He-core mass of $\sim$6 M$_{\odot}$, and thus a massive progenitor. For a Type IIP, SN\,2023ufx produced an unusually high amount of nickel ($^{56}$Ni) $\sim$0.14 $\pm$ 0.02 M$_{\odot}$, during the explosion.  We find that the short plateau duration in SN\,2023ufx can be explained with the presence of a small hydrogen envelope (M$_\mathrm{H_\mathrm{env}}$ $\simeq$1.2 M$_{\odot}$), suggesting partial stripping of the progenitor. About $\simeq$0.09 M$_{\odot}$ of CSM through mass loss from late-time stellar evolution of the progenitor is needed to fit the early time ($\lesssim$10 days) pseudo-bolometric light curve. Nebular line diagnostics of broad and multi-peak components of [O~I]~$\lambda\lambda$6300,\,6364, H$\alpha$, and [Ca II]~$\lambda \lambda$7291,\,7323 suggest that the explosion of SN\,2023ufx could be inherently asymmetric, preferentially ejecting material along our line-of-sight.
\end{abstract}

\section{Introduction} \label{sec:1}

Most massive stars ($\gtrsim$8 M$_{\odot}$) evolve rapidly, culminating in a cataclysmic core-collapse supernova (CCSN) explosion. Among these CCSNe, hydrogen-rich Type II SNe (SNe\,II) are the most commonly observed \citep[e.g.,][]{Li11, Smith11, Shivvers17}. The amount of hydrogen at core-collapse likely results in the diversity of observed subtypes of SNe\,II \citep[see][for a review]{Arcavi17}. Based on the shape of their light curves, \cite{Barbon_Ciatti_Rosino79} first proposed a division of SNe\,II into Type IIP (SNe\,IIP) and Type IIL (SNe\,IIL). 

In the SNe\,IIP subtype, the SN luminosity generally plateaus for $\sim$100 days after maximum as the ionized hydrogen in the envelope recombines \citep[e.g.,][]{Popov93, Kasen_Woosley09, Dessart_Hillier10, Goldberg19}. The subtype SNe\,IIL on the other hand shows a linear decline in luminosity after a rapid rise to peak luminosity \citep{Barbon_Ciatti_Rosino79, Filippenko97}. The stripped-envelope (SE) subtype SNe\,IIb retains only a few percent of the initial hydrogen envelope. Spectroscopically they show clear evidence for hydrogen and helium in the initial phases, but later the hydrogen lines become weak or absent \citep[e.g.,][]{Filippenko_Matheson_ho93}.

While SNe\,IIP and IIL were originally suggested to be two distinct classes \citep[e.g.,][]{Arcavi12, Faran14a, Faran14b}, larger statistical studies of IIP and IIL populations at optical wavelengths have supported a more continuous distribution of properties \citep{Patat94, Anderson14, Sanders15, Galbany16b, Valenti15, Valenti16}. SNe\,IIP and IIL show a continuous range of spectroscopic properties in the optical regime \citep[e.g.,][]{Gutierrez17_first, Gutierrez17_second}, but at near-infrared (NIR) wavelengths, a strong dichotomy has been reported between properties of SNe\,IIP and IIL suggesting possible differences in their ambient environments \citep{Davis19}. 

Hydrogen-rich SNe\,II that show narrow H emission lines indicating the presence of strong circumstellar material (CSM) interaction are classified as IIn \citep[see][for a review]{Schlegel90, Smith17}. Even in the absence of such narrow emission lines, hydrodynamic modeling of other SNe\,II (IIP/L) have suggested that CSM interaction might play a key role at early times \citep{Morozova17, Morozova18, Forster18, Moriya23}. They showed that SNe\,IIP from red supergiants (RSGs) with additional CSM around them could produce SNe\,IIL. Thus, SNe\,II with a rapid (IIL-like) early decline in their luminosity could be a direct probe into the immediate mass-loss history of the progenitor before explosion.

Pre-explosion imaging of the progenitors of SNe\,IIP have been identified as RSGs with a zero age main sequence (ZAMS) mass of $\sim$8 -- 17 M$_{\odot}$ \citep{Van_Dyk03, Smartt09, Smartt15, Van_Dyk17}. However, theoretical evolutionary codes predict the progenitors of SNe\,IIP can have initial masses $\lesssim$30 M$_{\odot}$ \citep[e.g.,][]{Heger03, Ekstrom12} and the observed RSG population in our Galactic neighborhood is expected to have a ZAMS mass range of $\sim$9--25 M$_{\odot}$ \citep[e.g.,][]{Levesque06, Gordon16}. This apparent discrepancy between progenitor masses of SNe\,II and our understanding of RSG evolution is often dubbed the RSG problem in stellar evolution \citep{Smartt15}. Due to uncertainties in late stage evolution of massive stars and the small sample size of SNe studied, the statistical significance of the RSG problem is highly debated and still is an open question \citep[e.g.,][]{Walmswell_Eldridge12, Eldridge13, Kochanek14, Meynet15, Sukhbold16, Adams17, Davies_Beasor18, Kochanek20, Davies_Beasor20, beasor24}. 

As plateau length is correlated with the duration of hydrogen recombination, analytical and numerical simulations of SNe\,IIP light curves have predicted a continuous scaling of photospheric plateau duration \citep{Popov93, Kasen_Woosley09, Sukhbold16, Hiramatsu21}, though there are some degeneracies with explosion energies and progenitor radii.  Short plateau SNe (SPSNe) with a plateau duration in the order of tens of days are rarely observed \citep[e.g.,][]{Hiramatsu21, Teja22, Teja23}. The origin of luminous SPSNe with a small hydrogen envelope (H-envelope) have been suggested to be due to partial stripping of a massive progenitor (M$_\mathrm{ZAMS}$ $\gtrsim$18 M$_{\odot}$) based on a small sample size \citep{Hiramatsu21, Martinez22}. The recent well-observed event SN~2023ixf appears to have arisen from a partially stripped and  similarly high-mass RSG progenitor as well \citep{hsu24}.

In this paper we present optical photometry and spectroscopy and NIR spectroscopy  of SN\,2023ufx, a SN\,II with an exceptionally short plateau. Recently, SN\,2023ufx was also studied by \cite{Tucker24}, where several unique aspects of its photometric and spectroscopic evolution were discussed in the context of massive star evolution. We will compare our interpretations with these results in the appropriate sections.

In Section \ref{sec:2}, we describe the discovery and observations, and in Section \ref{sec:3}, we estimate the interstellar extinction. We present the photometric and spectroscopic evolution of SN\,2023ufx in Section \ref{sec:4} and Section \ref{sec:5}, respectively. We discuss comparisons of our observations with synthetic models to infer progenitor properties in Section \ref{sec:6}. The proposed explosion asymmetry in SN\,2023ufx is discussed in Section \ref{sec:7} and we present our summary and conclusions in Section \ref{sec:8}.

\section{Discovery and Observations}  \label{sec:2}

SN\,2023ufx was discovered by the Asteroid Terrestrial-impact Last Alert System \citep[ATLAS;][]{Tonry18a, Smith20} on 2023 October 06 at 13:55:52 \citep[MJD = 60224.1;][]{Tonry23}. Throughout this work, all dates / times are reported in the Coordinated Universal Time (UTC) standard. 
The closest available non-detection before discovery is from the Zwicky Transient Facility \citep[ZTF;][]{Bellm19, Graham19, Masci23} on 2023 October 05 at 11:12:46 (MJD = 60222.5). Given a $\sim$1-day non-detection constraint, we adopt the explosion epoch (t$_{0}$) to be the mid-point between these two epochs at MJD = 60223.3 $\pm$ 0.5 (2023 October 06), where the uncertainties cover the full range of possible explosion epochs.

SN\,2023ufx was initially considered a fast evolving blue optical transient \citep[][]{Srivastav23, Tucker23}. However, Gemini GMOS-N spectroscopy classified it as a young Type IIP supernova at a redshift of $z$ = 0.0146 $\pm$ 0.0007 \citep[][]{Chrimes23}. We adopt this redshift in our work, which corresponds to a distance of $\sim$63.2 $\pm$ 3.1 Mpc \citep[based on the cosmological calculator\footnote{\url{https://astro.ucla.edu/~wright/CosmoCalc.html}} as described in][]{Wright06} and a distance modulus of 34.01 $\pm$ 0.11 assuming $H_{0}$ = 70 km s$^{-1}$ Mpc$^{-1}$, $\Omega_\mathrm{m}$ = 0.3, and $\Omega_\mathrm{vac}$ = 0.7 \footnote{The estimated distances will have significant uncertainties due to systematics in the adopted cosmology}. 

SN\,2023ufx was identified to be associated with the faint host galaxy SDSS J082451.43+211743.3 ($g$ = 20.86 $\pm$ 0.04 mag, $r$ = 20.62 $\pm$ 0.07 mag, $i$ = 20.86 $\pm$ 0.12 mag), exploding at a distance of $\approx$ 2.7 arcsec ($\sim$ 0.8 kpc) from its nucleus \citep{Tucker24}. These $gri$ magnitudes can be converted to $V$ = 20.72 $\pm$ 0.08 mag, based on empirical transformation relations described in \cite{Lupton05}\footnote{\url{https://classic.sdss.org/dr4/algorithms/sdssUBVRITransform.php}}. For our assumed distance of $\sim$63.2 $\pm$ 3.1 Mpc, this implies a host luminosity of log($L$/$L_{\odot}$) = 7.2 $\pm$ 0.1 dex, with $M_{V,\odot}$ = 4.84 as the absolute $V$ magnitude of the Sun. Adopting the luminosity-metallicity relationship of \cite{Kirby13}, we estimate [Fe/H] = -1.32 $\pm$ 0.19. Thus, the host metallicty of SN\,2023ufx can be expressed as log($Z$/$Z_{\odot}$) = -1.32 $\pm$ 0.19 dex. Such a low host-galaxy metallicity for SN\,2023ufx (as a SN IIP) is also generally consistent with independent estimate of log($Z$/$Z_{\odot}$) $<$ -0.9 dex by \cite{Tucker24} using a Keck Cosmic Web Imager \citep[KCWI;][]{Morrissey18} spectrum of SDSS J082451.43+211743.3. 


\subsection{Photometric Observations} \label{sec:2.1}
Shortly after discovery, we performed high-cadence photometric ($U$, $B$, $g$, $V$, $r$, and $i$) follow-up of SN\,2023ufx with the worldwide Las Cumbres Observatory network of 1.0\,m robotic telescopes  \citep{Brown13}. These observations were triggered through the Global Supernova Project \citep{Howell19}. The images were reduced using the PyRAF-based photometric reduction pipeline, \verb |lcogtsnpipe|\footnote{\url{https://github.com/LCOGT/lcogtsnpipe}} \citep{Valenti16}. The pipeline calculates the instrumental magnitudes using a standard point-spread function (PSF) fitting technique. Apparent magnitudes for $g$, $r$, and $i$ images were calibrated with the APASS catalog \citep{Henden16}. The $U$, $B$, and $V$ apparent magnitudes were calibrated with a Landolt catalog \citep{Landolt92} constructed using standard fields observed with the same telescope and night combination as the SN observations. Since background contamination from host-galaxy is minimal (faint host and significant offset from SN; see Section \ref{sec:2}), the PSF photometry was obtained directly from the unsubtracted images. 

We used the ATLAS forced photometry service \citep{Tonry18a, Smith20} to obtain photometry in two filters cyan ($c$) and orange ($o$). Additionally we obtained ZTF photometry data ($g$ and $r$) from the Automatic Learning for the Rapid Classification of Events (ALeRCE) broker \citep{Forster21}. 

SN~2023ufx was also observed by the Neil Gehrels \textit{Swift} Observatory \citep{Gehrels04}. The UVOT images were reduced using the High-Energy Astrophysics software (HEASoft\footnote{\url{https://heasarc.gsfc.nasa.gov/docs/software/heasoft/}}). The source region is centered at the position of the SN with an aperture size of 5$\arcsec$ and the background is measured from a region without contamination from other stars with an aperture size of 5$\arcsec$. We use a larger source aperture size than recommended\footnote{\url{https://swift.gsfc.nasa.gov/analysis/threads/uvot_thread_aperture.html}} as the PSFs in several images were not good due to Swift's temporary gyroscope related issues in October 2023. Zero-points for photometry were chosen from \cite{Breeveld10} with time-dependent sensitivity corrections updated in 2020. 

\subsection{Spectroscopic Observations} \label{sec:2.2}
We performed several optical and NIR spectral observations of SN\,2023ufx starting on 2023 October 13 ($\sim$6 days after discovery) and continued through 2024 May 04 ($\sim$211 days after explosion). The observation log associated with all optical and NIR spectra used in this work are presented in Table \ref{tab:optical_nir_spectralog} 

\subsubsection{Optical Spectroscopy} 

We obtained high-cadence low-resolution optical spectra using the FLOYDS spectrographs mounted on the 2.0\,m Faulkes Telescope North (FTN) at Haleakala (USA) and the identical 2.0\,m Faulkes Telescope South (FTS) at Siding Spring (Australia) which are part of the Las Cumbres Observatory network. The observations were performed with a 2.0-arcsec wide slit placed on the target at the parallactic angle. We extracted, reduced, and calibrated the 1D spectra using the standard FLOYDS reduction pipeline as described in \cite{Valenti14}.

We obtained 5 spectra with the Low-Resolution Imaging Spectrometer \citep[LRIS;][]{Oke95} mounted on the Keck~I telescope. The observations were performed with a 1.0-arcsec wide slit and the data were reduced utilizing standard procedures in the fully-automated reduction pipeline for LRIS longslit spectra, \verb |LPipe| \citep{Perley19}. 

An optical spectrum was obtained with the Multi-Object Double Spectrographs  \citep[MODS]{2010SPIE.7735E..0AP} on the twin 8.4\,m LBT at Mount Graham International Observatory. These spectra were reduced using standard techniques, including bias subtraction and flat fielding via the modsCCDRed package \citep{richard_pogge_2019_2647501}.  Cosmic ray rejection, local sky subtraction, and extraction of one-dimensional spectra were done using IRAF \citep{Tody86}, and flux calibration was done with standard-star observations taken on the same night at similar airmass.

We present a Southern Astrophysical Research Telescope (SOAR)  Goodman-RED spectrum taken on 2024-01-28 ($\sim$114 days after explosion). We reduced the spectrum using the Goodman HTS Pipeline\footnote{\url{https://soardocs.readthedocs.io/projects/goodman-pipeline/en/latest/}} and calibrated it to a spectrophotometric standard star observed on the same night using PyRAF \citep{PyRAF}.

We also present a spectrum from the Boller and Chivens Spectrograph (B\&C) on the University of Arizona's Bok 2.3\,m telescope taken on 2024-02-13 ($\sim$130 days after explosion). This spectrum was reduced using standard IRAF routines \citep{Tody86}.

We obtained three optical spectra with the Kast double spectrograph \citep{millerstone94} mounted on the Shane 3\,m telescope at Lick Observatory. A 2.0-arcsec slit was used with the slit oriented at the parallactic angle. The data were reduced following standard techniques for CCD processing and spectrum extraction \citep{Silverman12} utilizing IRAF \citep{Tody86} routines and custom Python and IDL codes\footnote{\url{https://github.com/ishivvers/TheKastShiv}}. The spectra were flux calibrated and telluric corrected using observations of appropriate spectrophotometric standard stars observed on the same night, at similar airmasses, and with an identical instrument configuration.

A nebular spectrum of SN\,2023ufx was taken with the Blue arm of the Hobby-Eberly Telescope Low Resolution Spectrograph 2 (HET LRS2-B) on 2024-05-04 ($\sim$211 days after explosion).  The spectrum of the SN was extracted from the background-subtracted data cube produced by the Panacea\footnote{\url{https://github.com/grzeimann/Panacea}} pipeline. A detailed prescription on reducing LRS2 data can be found in \citet{Thomas22}.

\subsubsection{Near-Infrared Spectroscopy} \label{sec:2.2.2}
We observed SN\,2023ufx in a \textit{zJ}  spectroscopic setup on 2023-12-03 ($\sim$58 days after explosion) using the MMT and Magellan Infrared Spectrograph \citep[MMIRS;][]{mmirs} on the 6.5\,m MMT located on Mt.\,Hopkins in Arizona. The spectrum was taken using a 1.0-arcsec longslit. The data were manually reduced using the MMIRS pipeline \citep{mmirspipe}, then the 1-D spectral outputs were telluric and absolute flux corrected following the method described in \citet{Vacca03} with the \verb|XTELLCOR_GENERAL| tool \citep[part of Spextool package]{Cushing04} using a standard A0V star observed at similar airmass and time. 

\begin{figure}
    \centering
    \includegraphics[width=0.48\textwidth] {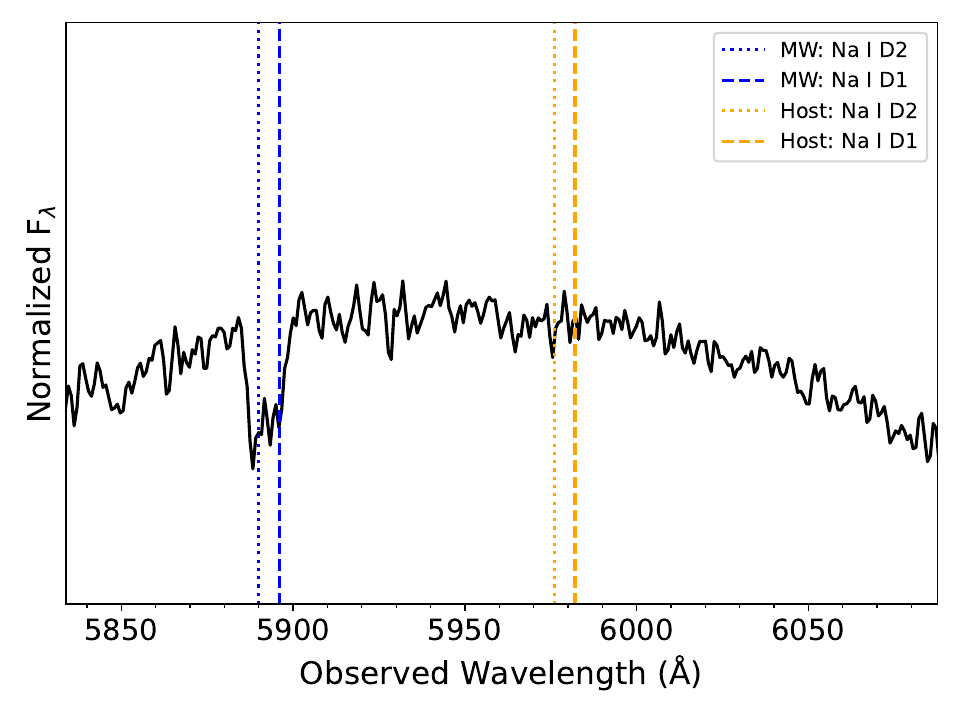}
    \caption{Red-side MODS LBT spectrum of SN\,2023ufx on 2023-11-09. Na I D2 and Na I D1 expected from the host galaxy (orange) and Milky Way (blue) are marked. There is no discernible absorption due to the host galaxy.} 
    \label{fig:NaID}
\end{figure}

\begin{figure*}
    \centering
    \includegraphics[scale=0.6] {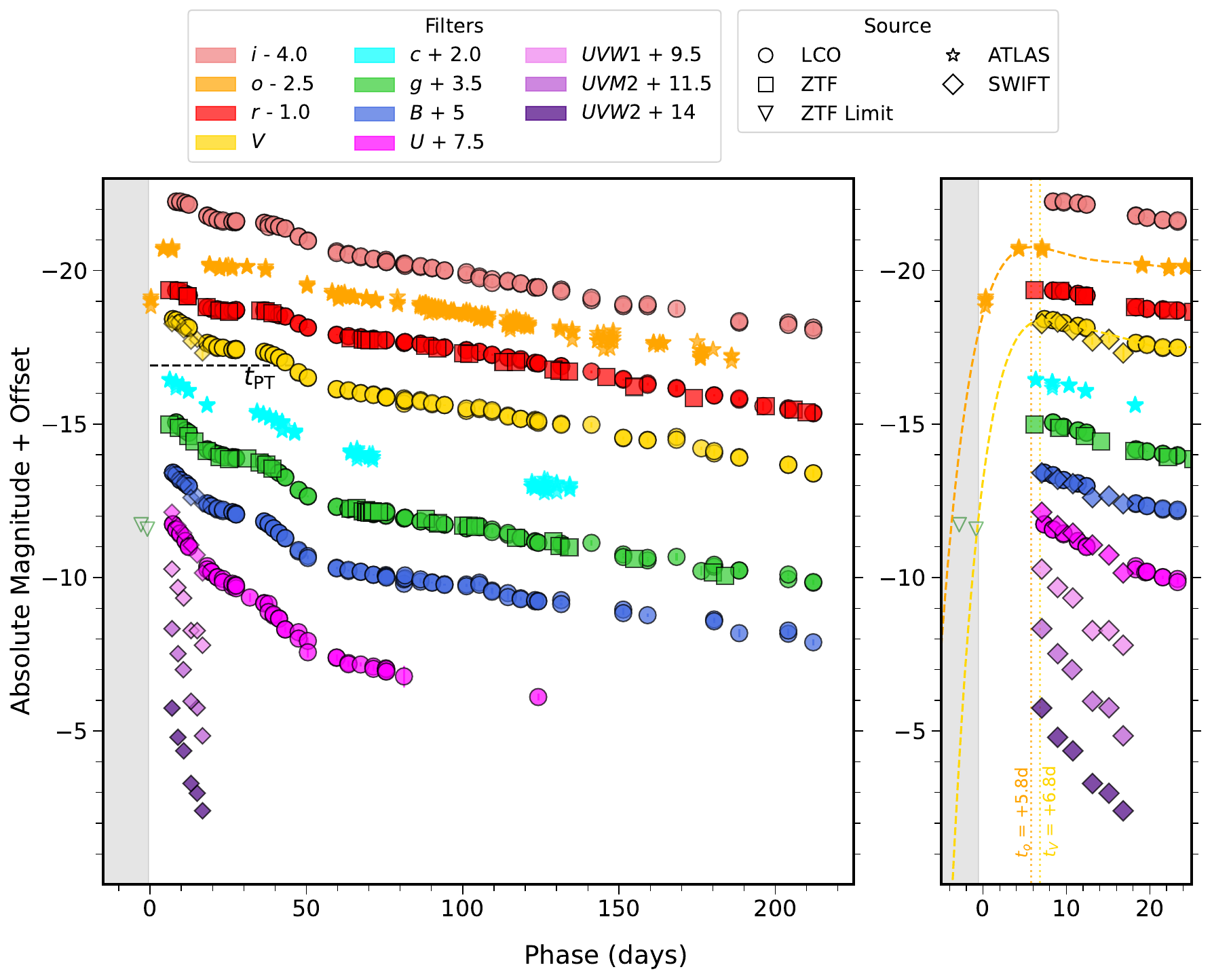}
    \caption{\textit{(Left):} Multi-band extinction-corrected light curves of SN\,2023ufx from Las Cumbres Observatory, ZTF, ATLAS, and Swift data with respect to epoch of explosion (t$_{0}$) on 2023-10-06. The SN\,2023ufx $V$ band plateau length ($t_\mathrm{PT}$) as described and measured in Section \ref{sec:4.2} is marked.  \textit{(Right):} Zoom in showing the first ATLAS ($o$) detection of SN\,2023ufx and last known non-detection from ZTF ($g$), along with other bands. Epochs of peak brightness in $o$ (t$_{o}$) and $V$ (t$_{V}$) are within $\sim$1 day of each other and are marked (dotted lines). Best-fit sixth order polynomials to estimate peak magnitude in $o$ and $V$ are plotted (dashed curves). Between last ZTF ($g$) non-detection and first $o$ detection ($\sim$0.5 days), the brightness jumped by $\sim$2 mag. Gray band in both panels indicate the time before SN detection.}
    \label{fig:photometry}

\end{figure*}

We observed SN\,2023ufx with the TripleSpec spectrograph \citep{Schlawin14} on the SOAR telescope on 2023-12-18 ($\sim$73 days after explosion). The spectrum was taken in the cross-dispersed mode using a 1.1-arcsec longslit and was reduced following the TripleSpec specific modification \citep{Kirkpatrick11}\footnote{\url{https://noirlab.edu/science/index.php/observing-noirlab/observing-ctio/observing-soar/data-reduction/triplespec-data}} of the reduction software, \verb|Spextool| \citep{Cushing04}. The standard A0V star observed at similar airmass adjacent to the science target was used for telluric correction following the prescription in \cite{Vacca03}.

We obtained a NIR spectrum of SN\,2023ufx on 2023-12-30 ($\sim$85 days after explosion), with the Near-InfraRed Echellette Spectrometer \citep[NIRES;][]{Wilson04} on Keck II telescope. The observations were performed with NIRES fixed single slit (0.55-arcsec $\times$ 18-arcsec) and data were reduced using the semi-automated \verb|Python|-based open-source facility spectroscopic reduction software \verb|Pypeit| \footnote{\url{https://pypeit.readthedocs.io/en/release/tutorials/nires_howto.html}}. \verb|Pypeit| automatically identifies the trace in the 2D frames and performs spectral 1D extractions. A sensitivity function is computed by comparing observations of the A0V standard to the model spectrum and the telluric absorption spectrum at Maunakea constructed using the \verb|Telfit| code \citep{Gullikson14}. This sensitivity function is then used to flux calibrate and co-add science observations. 

\section{Extinction} \label{sec:3}


The equivalent widths (EWs) of Na I D absorption lines can be empirically related to the overall reddening toward a SN as it is generally a good tracer of gas and dust \citep[e.g.,][]{poznanski+12}. We measured the reddening along the line-of-sight toward SN 2023ufx by considering contributions from Milky Way and the host galaxy. In Figure \ref{fig:NaID}, we present a medium-resolution (R $\sim$2000) red-side MODS spectrum obtained on November 9, 2023. 

To estimate the EWs of Na I D line absorption due to Milky Way (MW), we continuum-normalize the observed spectrum and fit the blended Na I D2 and Na I D1 absorption due to MW with a single Gaussian. We estimate a total EW of Na I D from MW to be 0.52 $\pm$ 0.07 \AA. From the relation between EW and $E(B-V)$ (see equation 9 of \citealt{poznanski+12}), we estimate a corresponding total reddening of $E(B-V)_\mathrm{MW}$ = 0.049 $\pm$ 0.014 mag. There are no discernible Na I D absorption dips associated with the host galaxy at observed wavelengths, $\lambda$5976 D2 and $\lambda$5982 D1 (Figure \ref{fig:NaID}). We estimate an upper limit on the EW of Na I D from the host to be $<$ 0.04 \AA. This corresponds to $E(B-V)_\mathrm{host}$ $<$ 0.01 mag. As this upper limit is comparable to the uncertainty in $E(B-V)_\mathrm{MW}$, we assume $E(B-V)_\mathrm{host}$ to be negligible.

Reddening due to MW in the direction of SN 2023ufx based on the dust maps of \cite{Schafly11} is $E(B-V)_\mathrm{MW}$ = 0.042 $\pm$0.0013 mag, which is consistent with the reddening estimate from our EW measurement of Na I D lines. In this paper we adopt this latter value,  i.e., $E(B-V)_\mathrm{MW}$ = $E(B-V)_\mathrm{total}$ $\sim$0.04 mag, assuming the extinction law of \cite{cardelli89} with R$_\mathrm{v}$ = 3.1 for multi-band extinction corrections.

\section{Photometric Evolution} \label{sec:4}
The full extinction-corrected multi-band light curves of SN\,2023ufx are shown in Figure \ref{fig:photometry}. In the first $\sim$6 days after detection, only ATLAS $o$ data was obtained, making the rise time of SN\,2023ufx in other bands uncertain. To estimate peak brightness in $V$ we fit a window of 35 days around the earliest observed photometry using high-order polynomials. We find that the data are best-fit with a sixth-order polynomial (Figure \ref{fig:photometry}; right panel) and the measured peak brightness is $M_{V}$ = -18.42 $\pm$ 0.08 mag on MJD 60230.12, $\sim$7 days after explosion. Since ATLAS $o$ is the only filter where we catch the rise for SN\,2023ufx, we also estimate its magnitude and epoch of peak luminosity. Following the same fitting procedure as for $V$ we find that the peak brightness in $o$ is $M_{o}$ = -18.27 $\pm$ 0.02 mag on MJD 60229.1. This is generally consistent with our predicted epoch of peak magnitude in $V$ (given a $\sim$1 day uncertainty in the explosion epoch) and we can constrain the rise time to peak brightness in SN\,2023ufx to be $\lesssim$7 days.

\subsection{Early Rapid Decline} \label{sec:4.1}
Soon after the first Las Cumbres Observatory and ZTF observations, a rapid multi-band decline is observed, with faster decline in bluer bands. In order to understand where SN\,2023ufx lies compared to other SNe\,II in the literature, we quantify the rate of decline after maximum light.

\begin{figure}
    \centering
    \includegraphics[width=0.5\textwidth] {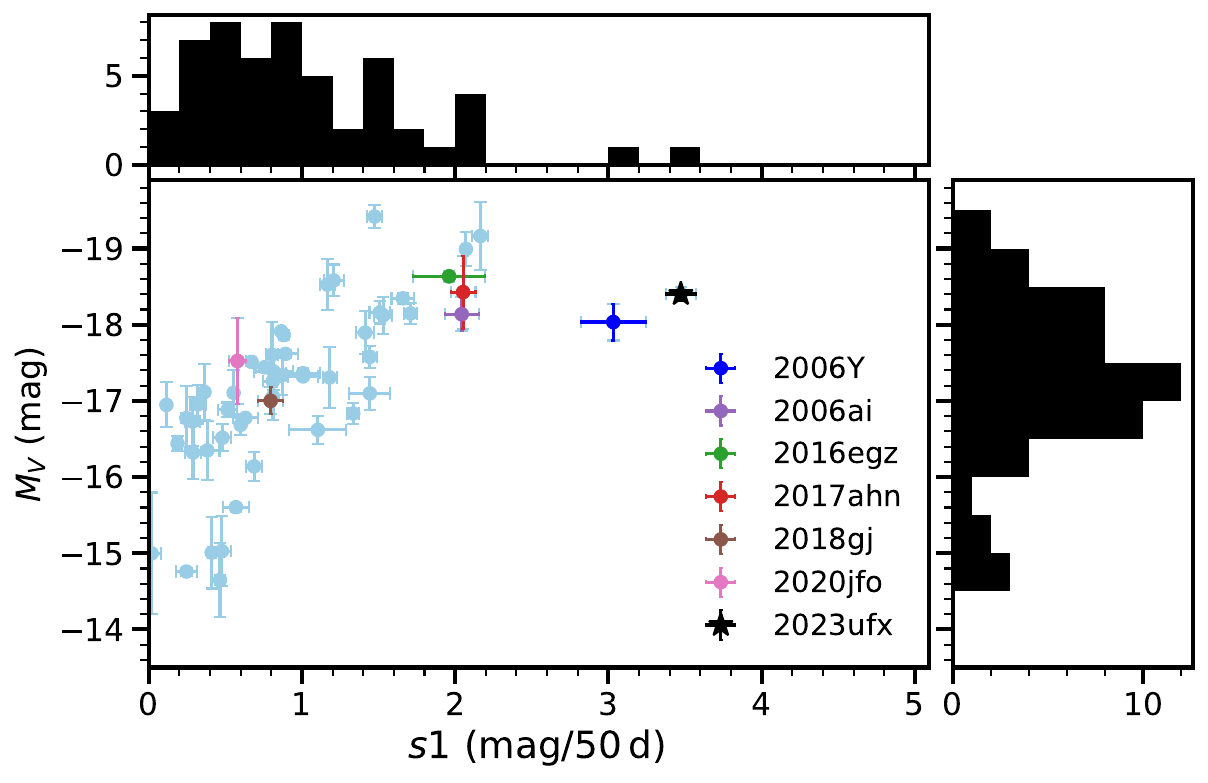}
    \caption{The peak magnitude, $M_{V}$ compared to early decline slope, $s1$ for SN\,2023ufx and a sample of SNe\,II from \cite{Anderson14} and \cite{Valenti16}, where there is a distinct slope at early times. Some other SPSNe are marked. SN\,2023ufx is both luminous and is one of the fastest declining SNe\,II at early times.}
    \label{fig:Mv_s1}
\end{figure}

Comprehensive statistical studies of decline rates in brightness have been shown to have a correlation with the peak absolute magnitude \citep[see][]{Li11, Anderson14, Galbany16b, Valenti16}. It has also been noted that some SNe\,II show a change in the decline rate around 10--20 days \citep{Barbon_Ciatti_Rosino79, Anderson14}. For these SNe, a single slope cannot account for the different decline rates until the end of hydrogen recombination. SN\,2023ufx has a significant slope change in all bands around day 20, so we follow the procedure in \cite{Valenti16} to measure the decline rate of the initial steeper slope of the light curve in $V$, $s1$. To be consistent with definitions in the literature, we present $s1$ in units of mag / 50 days as described in \cite{Anderson14}. 

For SN\,2023ufx, the best-fit slope for early decline is $s1$ = 3.47 $\pm$ 0.09 mag / 50 days. This is the fastest early decline observed among all SNe\,II samples in \cite{Anderson14} and \cite{Valenti16}, where there is a discernible early decline before the plateau phase. Along with a high $M_{V}$, this puts SN\,2023ufx in a previously unexplored parameter space. We present this comparison in Figure \ref{fig:Mv_s1}, where we combined the sample data in \cite{Anderson14} and \cite{Valenti16}.  SN\,2018ivc is another unique SN II, where four clear slope changes during the decline were observed in the first 30 days of evolution, though it did not have a single plateau like a SN IIP \citep{Bostroem20, Reguitti24}. We mark other known SNe\,II with a single short plateau SNe 2006Y, 2006ai, 2016egz \citep{Hiramatsu21}, 2018gj \citep{Teja22}, and 2020jfo \citep{Teja22} in the plot. We also mark the more IIL-like 2017ahn \citep{Tartaglia21}. All the high peak luminosity SPSNe (especially SN\,2006Y and SN\,2023ufx) have faster than average (SNe\,II) decline rates. We discuss the potential origin of such a rapid early decline further in Section \ref{sec:6.1}.  

\subsection{Short-Plateau} 

After the rapid decline in the first $\sim$20 days, SN\,2023ufx settles on an extremely short plateau (of $\sim$20 days), before a significant drop in brightness as the hydrogen recombination phase ends. While other significantly short plateau SNe (with $\lesssim$50 days on the plateau) have been discussed previously in the literature \citep{Hiramatsu21}, SN\,2023ufx is the shortest yet. 

\begin{figure}
    \centering
    \includegraphics[width=0.5\textwidth] {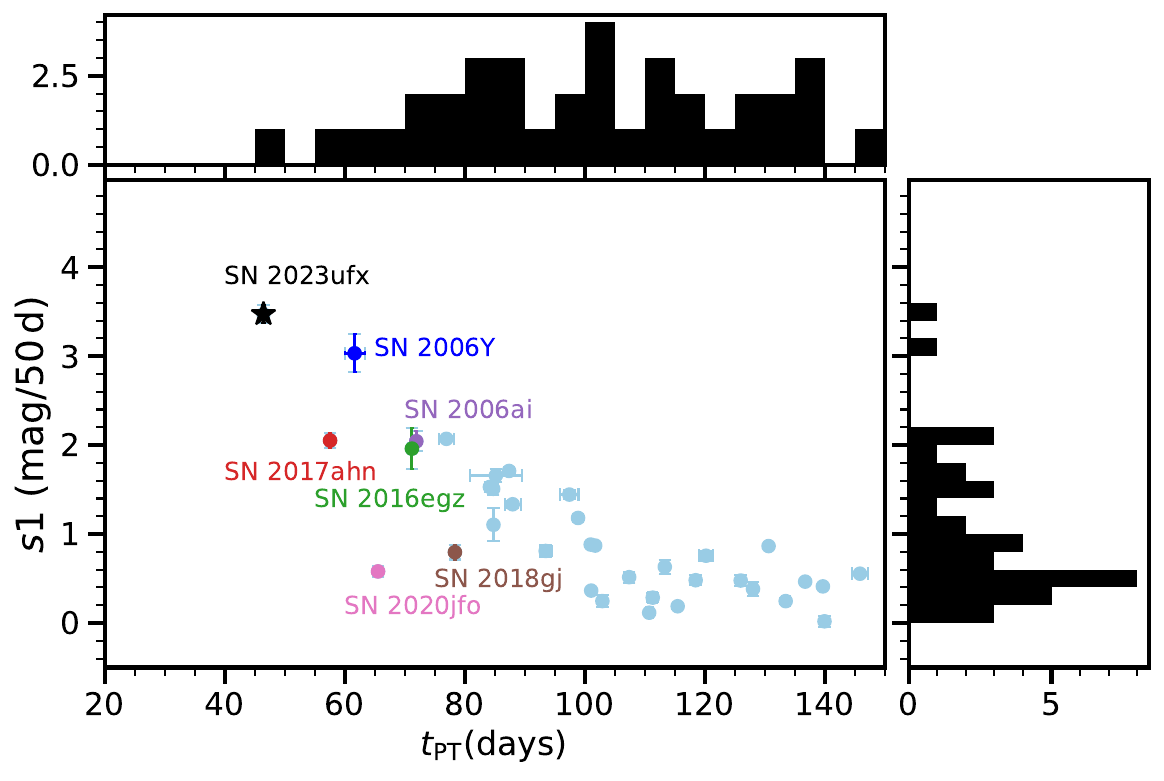}
    \caption{The early decline rate, s1 is plotted against the plateau length $t_\mathrm{PT}$ for a sample of SNe\,IIP including SN\,2023ufx and some other SPSNe. SN\,2023ufx has the shortest plateau length among all SNe\,IIP in literature.}
    \label{fig:s1_vs_tpt}
\end{figure}

During the fall from the plateau phase, the observed light curve can be characterized by a Fermi-Dirac phenomenological formulation \citep{Olivares10, Valenti16, Dong21}:
\begin{equation}
y(t) = \frac{-a0}{1+e^{(t-t_\mathrm{PT})/w_{0}}} + (p0 \times (t-t_\mathrm{PT})) + m0
    \label{eqn1}
\end{equation}
where $a0$ indicates the depth of the drop, $t_\mathrm{PT}$ represents the ``length of plateau'' as described in \cite{Valenti16}, and $w0$ inversely indicates the slope of the post-plateau light curve phase before the radioactive tail phase begins. We fit the $V$-band light curve with this phenomenological model using Markov Chain Monte Carlo (MCMC) sampling with the Python package, \verb|emcee| \citep{ForemanHogg13}. The corresponding best-fit values are: $a0$ = 1.16 $\pm$ 0.08 mag;  $t_\mathrm{PT}$ = 46.5 $\pm$ 0.5 days; and $w0$ = 4.6 $\pm$ 0.6 days. We uniformly vary our priors for $a0$ (0.5 -- 2.5 mag), $t_\mathrm{PT}$ (0 -- 200 days), and  $w0$ (1 -- 20 days). All other SNe in our sample are fit in the same way as described for SN\,2023ufx to ensure a consistent comparison of $t_\mathrm{PT}$. We find that SN\,2023ufx has the shortest $t_\mathrm{PT}$ among all SNe in our comparison database. We plot $s1$ with respect to $t_\mathrm{PT}$ in Figure \ref{fig:s1_vs_tpt}, where SN\,2023ufx and other SPSNe are marked.  From Figure \ref{fig:Mv_s1} and Figure \ref{fig:s1_vs_tpt} it is clear that luminous short-plateau SNe explore a different parameter space in their early-phase light curve evolution compared to  a more typical SN\,II. 

\label{sec:4.2}

\begin{figure*}
    \centering
    \includegraphics[scale=0.45] {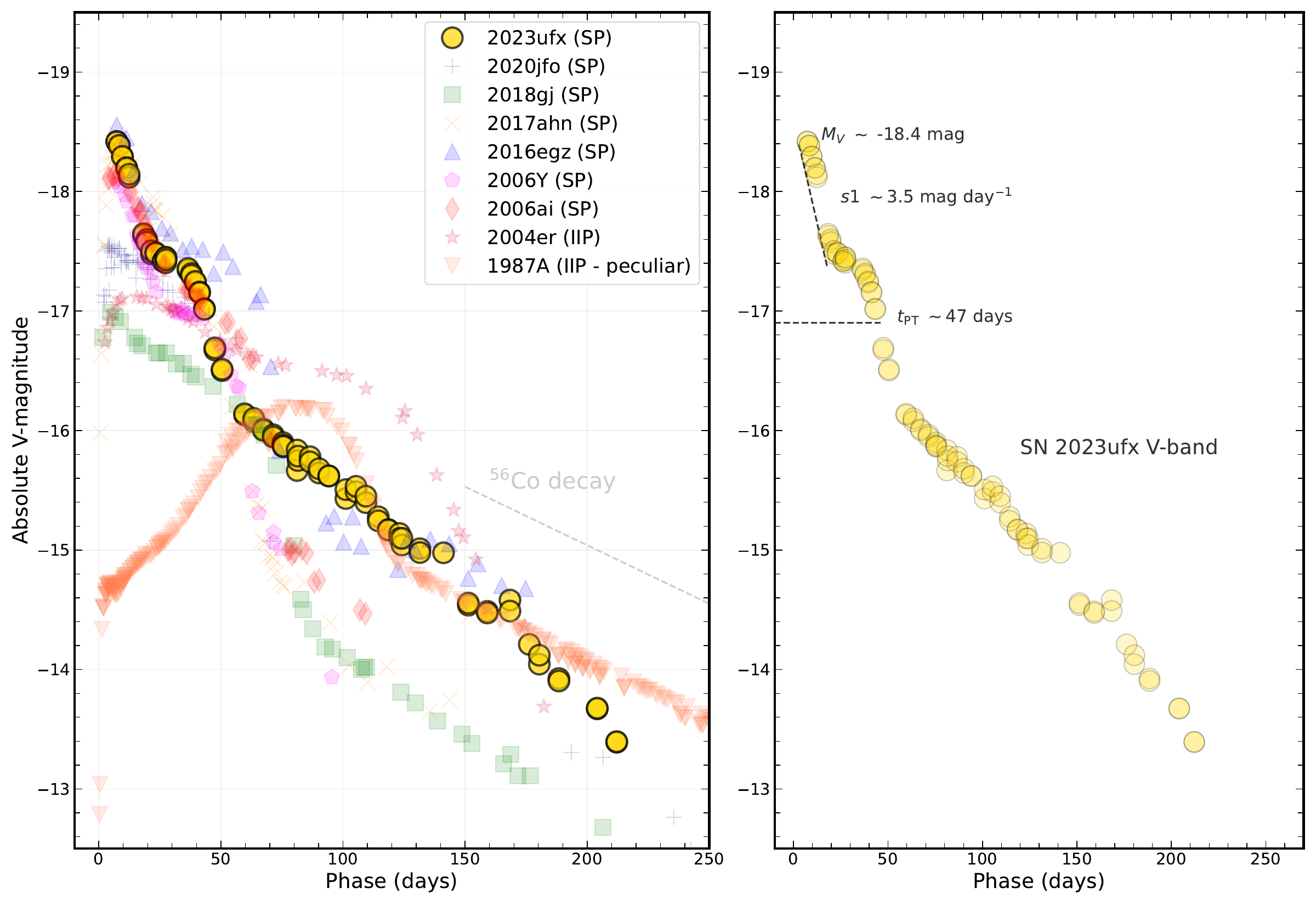}
    \caption{\textit{Left}: Absolute magnitude V-band light curve comparison between SPSNe (including SN\,2023ufx) and a more typical SN IIP (SN\,2004er) with a significantly longer plateau phase duration. The peculiar SN IIP 1987A along with the typical $^{56}$Co-decay rate are also plotted for comparison. SN\,2023ufx is among the most luminous SPSNe with a rapid early decline before the plateau phase. The references for data on other SNe used in this plot are as follows: SN 2020jfo \citep{Teja22}, SN 2018gj \citep{Teja23}, SN 2017ahn \citep{Tartaglia21}, SN 2016egz, SN 2006Y, SN 2006ai \citep{Hiramatsu21}, SN 2004er \citep{Anderson14}, and SN 1987A \citep{Menzies87, Catchpole87, Catchpole88, Richardson01, Tsvetkov04}. \textit{Right}: Absolute magnitude V-band light curve of SN\,2023ufx with the photometric parameters, $M_{V}$, $s1$, and $t_\mathrm{PT}$ as defined in Section \ref{sec:4.1} and Section \ref{sec:4.2} marked.}
    \label{fig:SPSNe_comparison}
\end{figure*}

\subsection{Nickel Mass} \label{sec:4.3}
\begin{figure}
    \centering
    \includegraphics[width=0.5\textwidth] {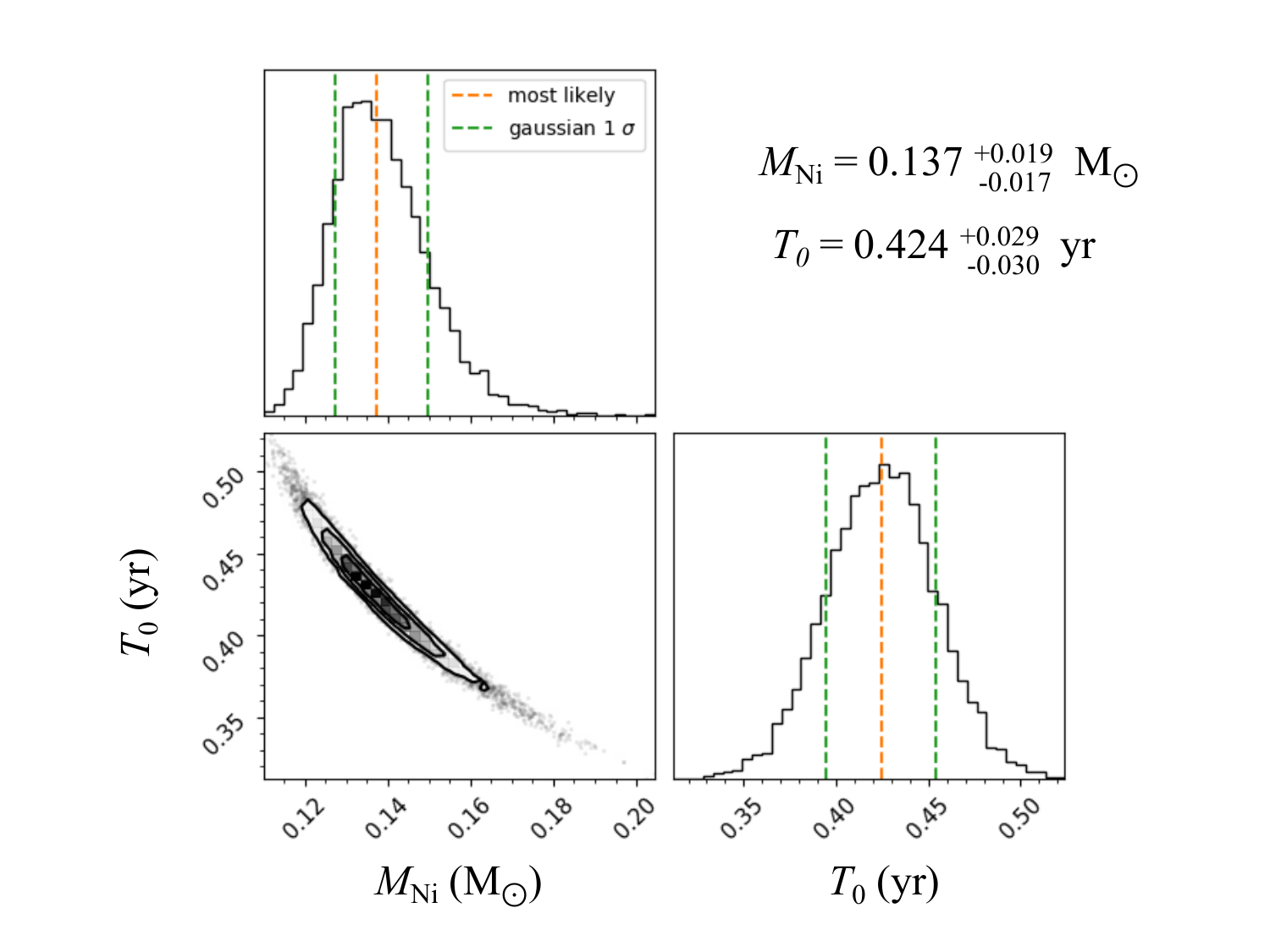}
    \caption{Posterior distribution of nickel mass, M$_\mathrm{Ni}$ and the characteristic trapping parameter, $T_{0}$ from MCMC sampling of the radioactive decay tail of SN\,2023ufx assuming incomplete trapping of $\gamma$-rays. The mean of the posterior distribution and 1-$\sigma$ uncertainties are marked.}
    \label{fig:MNi}
\end{figure}

\begin{figure*}
    \centering
    \includegraphics[width=0.8\textwidth] {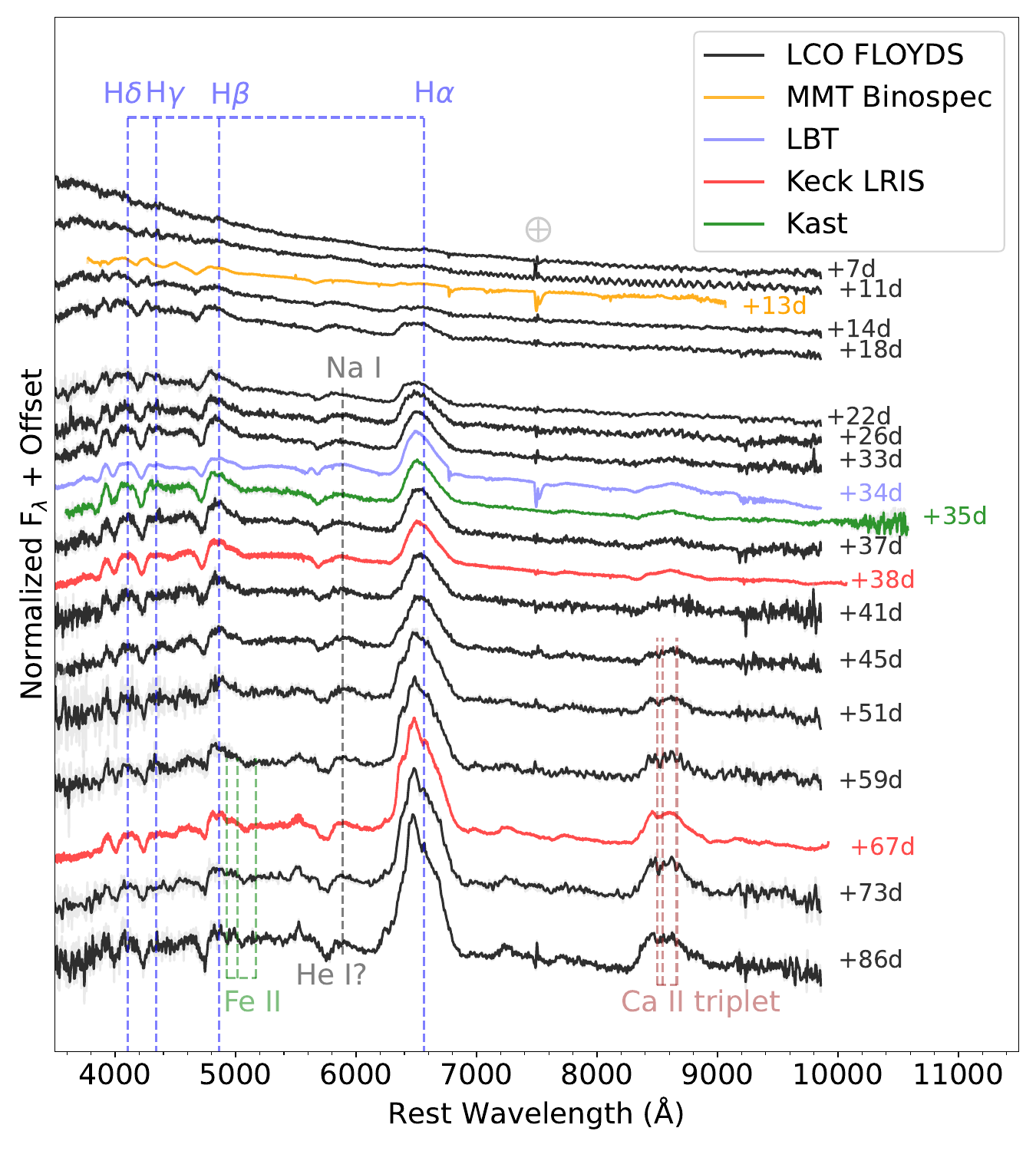}
    \caption{Evolution of photospheric optical spectra of SN\,2023ufx between 7 and 86 days. All spectra have been corrected for redshift and total reddening. A minimal smoothing has been performed on some of the spectra where the original un-smoothed data are in light gray. Identified spectral features are marked. }
    \label{fig:photospheric_spectra}
\end{figure*}

After the fall from plateau, the light curve of SN\,2023ufx starts on the radioactive tail phase, presumed to be powered primarily by the radioactive decay of $^{56}$Co $\rightarrow$ $^{56}$Fe. We compare the $V$-band light curves of the SPSNe discussed in this work and a typical SN II in the left panel of Figure \ref{fig:SPSNe_comparison}.  We find that SN\,2023ufx is most similar to other luminous SPSNe (SN\,2006Y, SN\,2006ai, and SN\,2016egz), though with an even shorter plateau length. In the right panel of Figure \ref{fig:SPSNe_comparison}, we indicate the photometric parameters discussed in Sections \ref{sec:4.1} and \ref{sec:4.2} on the absolute V-band light curve of SN\,2023ufx. We find that the brightness of SN\,2023ufx declines faster ($\sim$0.02 mag day$^{-1}$) than expected with complete gamma ray trapping ($\sim$0.0098 mag day$^{-1}$) at these times. 

The amount of $^{56}$Ni produced through explosive nucleosynthesis can be estimated from evolution of the radioactive tail of the optical light curve, during the optically thin nebular phase. The spontaneous radioactive decay processes produce $\gamma$-rays that are reprocesssed by the ejecta, radiating at optical wavelengths. If these $\gamma$-rays are completely trapped by the ejecta (i.e., the same trapping as SN\,1987A), the $^{56}$Ni mass at explosion can be estimated by scaling the pseudo-bolometric luminosity of the SN at nebular times to that of the extremely well characterized SN\,1987A (assuming the SN has the same SED as SN\,1987A) with the following equation \citep[see][and references therein]{Spiro14}. 
\begin{equation}
    M_\mathrm{Ni} = 0.075 M_{\odot} \times \frac{L_\mathrm{SN}(t)}{L_\mathrm{87A}(t)} 
    \label{eqn:1}
\end{equation}
where M$_\mathrm{Ni}$ is the synthesized $^{56}$Ni mass, $L_\mathrm{SN}(t)$ and  $L_\mathrm{87A}(t)$ are the pseudo-bolometric luminosities of the SN of interest and SN\,1987A at time $t$, constructed using a common set of filters. 
However, from a sample study of SNe\,II, \cite{Anderson14, Gutierrez17_second} showed that most fast-declining SNe show a radioactive tail decline faster than that expected from $^{56}$Co decay with complete trapping, suggesting an incomplete trapping of $\gamma$-rays. To account for the luminosity deficit due to incomplete trapping, we can modify Equation \ref{eqn:1} by simplifying the incomplete trapping terms in \cite{Clocchiatti_Wheeler97} as 
\begin{align}
M_\mathrm{Ni} = 0.075 M_{\odot} &\times \frac{1 - e^{-(530/t)^2}}{1 - e^{-(T_\mathrm{0}/t)^2}} \times \frac{L_\mathrm{SN}(t)}{L_\mathrm{87A}(t)} 
\label{eqn:2}
\end{align}
where, $T_\mathrm{0}$ is the characteristic trapping parameter that represents $\gamma$-ray escape time associated with the SN. The trapping parameter associated with SN\,1987A is fixed at 530 days \citep{Anders11}. As the decline rate of SN\,2023ufx is faster than typically expected from complete trapping like in SN\,1987A (see Figure \ref{fig:SPSNe_comparison}), we use Equation \ref{eqn:2} to estimate the mass of $^{56}$Ni. 

For estimating the pseudo-bolometric luminosity of SN\,2023ufx, we followed the method described in \cite{Valenti08}. We converted the observed magnitudes to flux at each band and numerically integrated using the Simpson's rule. We perform a two-parameter (M$_\mathrm{Ni}$ and $T_\mathrm{0}$) MCMC sampling based on Equation \ref{eqn:2}. For this, we compare the pseudo-bolometric luminosities of SN\,2023ufx and SN\,1987A in the common set of ${UBgVri}$ filters between days 125 and 225, when both objects are on their respective radioactive tails. Note that for SN\,2023ufx, the radioactive tail phase starts as early as $\sim$60 days (Figure \ref{fig:photometry}, Figure \ref{fig:SPSNe_comparison}), but at this time SN\,1987A is still in the photospheric phase (Figure \ref{fig:SPSNe_comparison}). For our sampling, we use a log likelihood function based on the standard $\chi^{2}$ formulation. 
Results from our MCMC sampling are presented as posterior distributions in Figure \ref{fig:MNi}. We find a best-fit M$_\mathrm{Ni}$ = 0.137 $^{+0.019}_{-0.017}$ M$_{\odot}$, where the systematic errors on M$_\mathrm{Ni}$ due to distance are included. The best-fit trapping parameter for SN\,2023ufx, $T_\mathrm{0}$ = 154.93 $^{+10.67}_{-10.89}$ is significantly smaller than that of SN\,1987A \citep[i.e., $T_\mathrm{0}$ = 530 days;][]{Anders11} as expected. 

The estimated M$_\mathrm{Ni}$ of SN\,2023ufx is significantly higher than the observed median nickel mass (M$_\mathrm{Ni}$ = 0.03 M$_{\odot}$) for several SNe\,II samples \citep[see][]{Anderson14, Valenti16, Rodriguez21}.  This is consistent with high nickel mass estimates for other luminous SPSNe \citep{Hiramatsu21}. One general caveat to note with this method of estimating M$_\mathrm{Ni}$ is the assumption that all luminosity at nebular phases is coming from radioactive decay. If there is additional powering due to interaction (as we discuss later in Sections \ref{sec:5.2}, \ref{sec:6.2}, and \ref{sec:7}), this M$_\mathrm{Ni}$ of SN\,2023ufx can be an overestimate. 

Hydrodynamic light curve models of SNe\,II find that as a RSG loses H-envelope mass, the plateau becomes shorter - but also declines more rapidly \citep{Hiramatsu21}. They find that SPSNe in particular may be the result of stars with very small H envelopes left at core-collapse and high M$_\mathrm{Ni}$ (M$_\mathrm{Ni}$ $\gtrsim$0.05 M$_{\odot}$). The high M$_\mathrm{Ni}$ observed in SN\,2023ufx, one of the shortest-plateau SN in literature is consistent with these results. We further discuss comparisons between hydrodynamic model grids in \cite{Hiramatsu21} and the pseudo-bolometric light curve of SN\,2023ufx in Section \ref{sec:6.1}

\begin{figure*}
    \centering
    \includegraphics[width=0.8\textwidth] {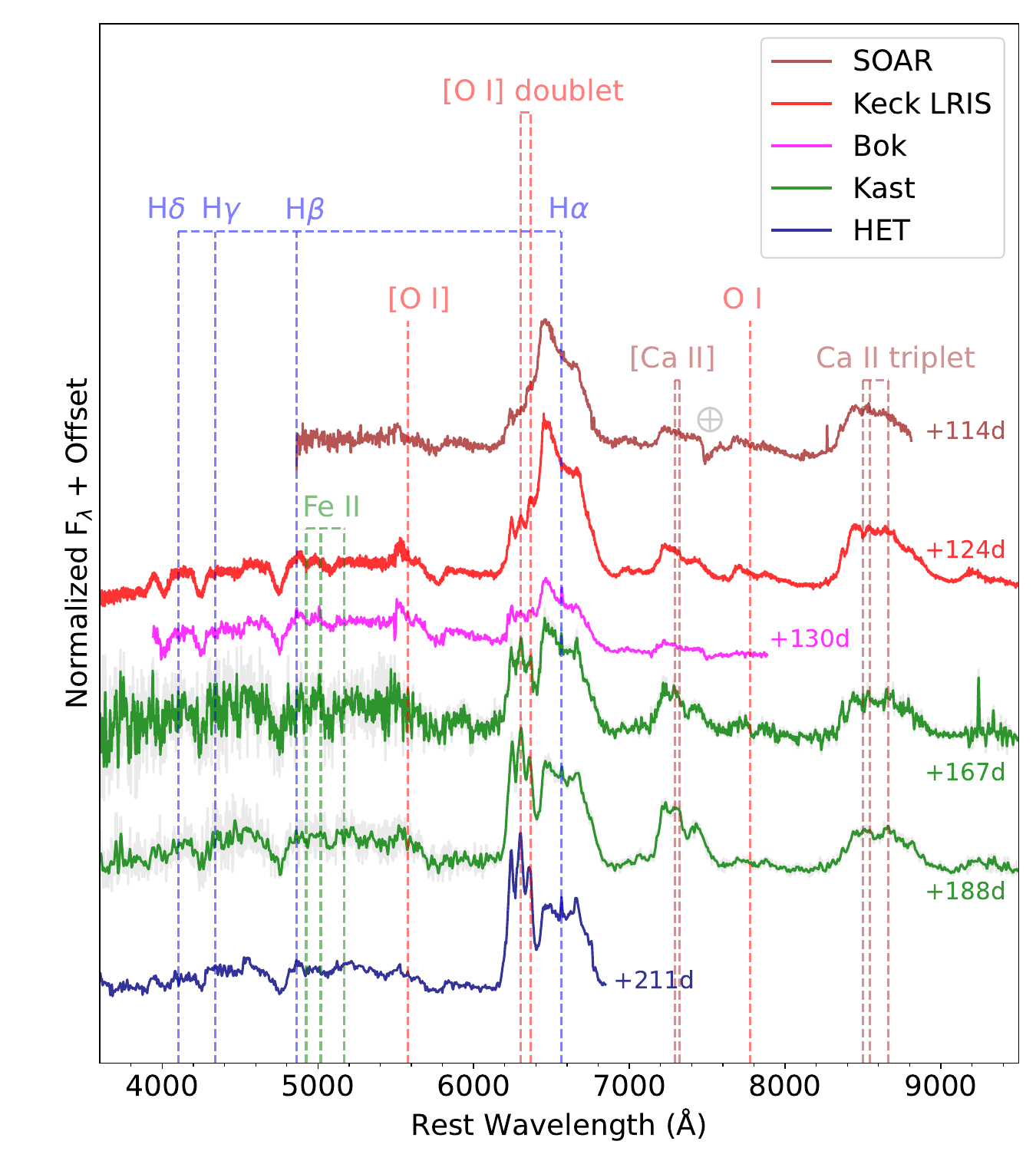}
    \caption{Evolution of late-time optical spectra of SN\,2023ufx between 114 and 211 days. All spectra have been corrected for redshift and total reddening. Identified spectral features are marked. High-velocity additional peaks are observed for [O~I] $\lambda \lambda$ 6300, 6364, H$\alpha$, and [Ca~II] $\lambda \lambda$ 7321, 7339 emission lines. A marginal smoothing has been performed on some of the spectra where the original un-smoothed data are in light gray.}
    \label{fig:nebular_spectra}
\end{figure*}

\begin{figure}

    \centering

    \includegraphics[width=0.5\textwidth] {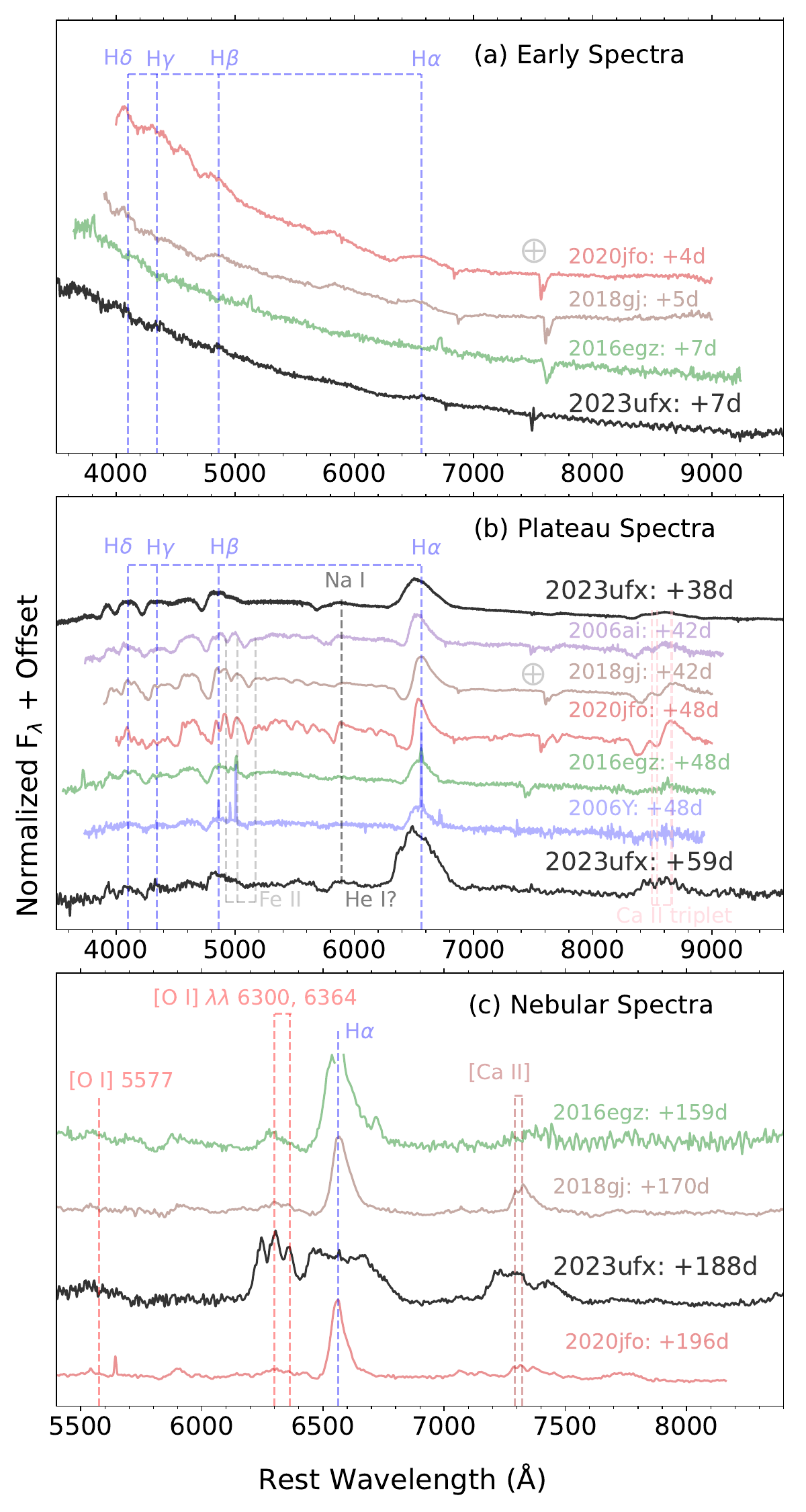}    
    \caption{Comparison of redshift- and extinction-corrected optical spectra between the SPSN sample in this work at a) early, b) plateau, and c) nebular phases. No early flash features are observed in these SPSNe. Fe~II absorption in SN\,2023ufx is significantly weaker than other SPSNe, even near the end of the plateau phase ($\sim$59 days). SN\,2023ufx has the lowest absorption to emission strength in H$\alpha$ among all the SPSNe sample. The nebular emission lines of O, H, and Ca in SN\,2023ufx are significantly different in width and morphology compared to other SPSNe. Reference for data: SN\,2006Y and SN\,2006ai \citep{Gutierrez17_first}; SN\,2016egz \citep{Hiramatsu21}; SN\,2018gj \citep{Teja23}; SN\,2020jfo \citep{Teja22}. }
    \label{fig:optical_spectra_comparison}
\end{figure}

\begin{figure}
    \centering
    \includegraphics[width=0.5\textwidth] {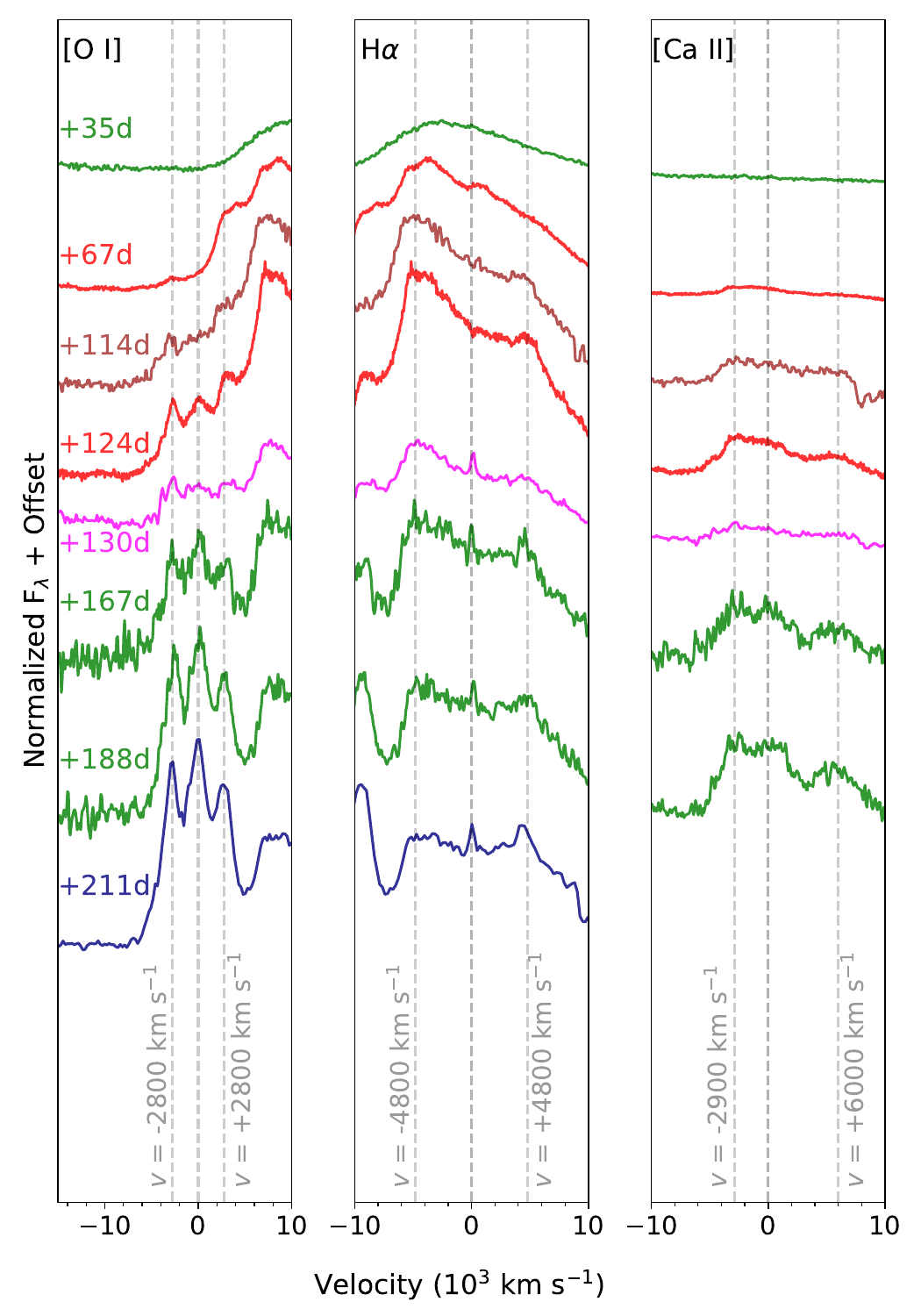}
    \caption{Evolution of the velocity profiles of [O~I]~$\lambda\lambda$6300,\,6364 doublet (left), H$\alpha$ (middle), and [Ca~II]~$\lambda\lambda$7321,\,7339 doublet (right) at few representative epochs between photospheric and nebular phases. The zero velocities for [O~I]~$\lambda\lambda$6300,\,6364 doublet, H$\alpha$, and [Ca~II]~$\lambda\lambda$7321,\,7339 doublet are at 6300, 6564, and 7321 \AA\, respectively. The velocities associated with the observed peaks calculated relative to the corresponding zero velocities in each panel have been marked.}
    \label{fig:optical_velocity}
\end{figure}

\begin{figure}
    \centering
    \includegraphics[width=0.48\textwidth] {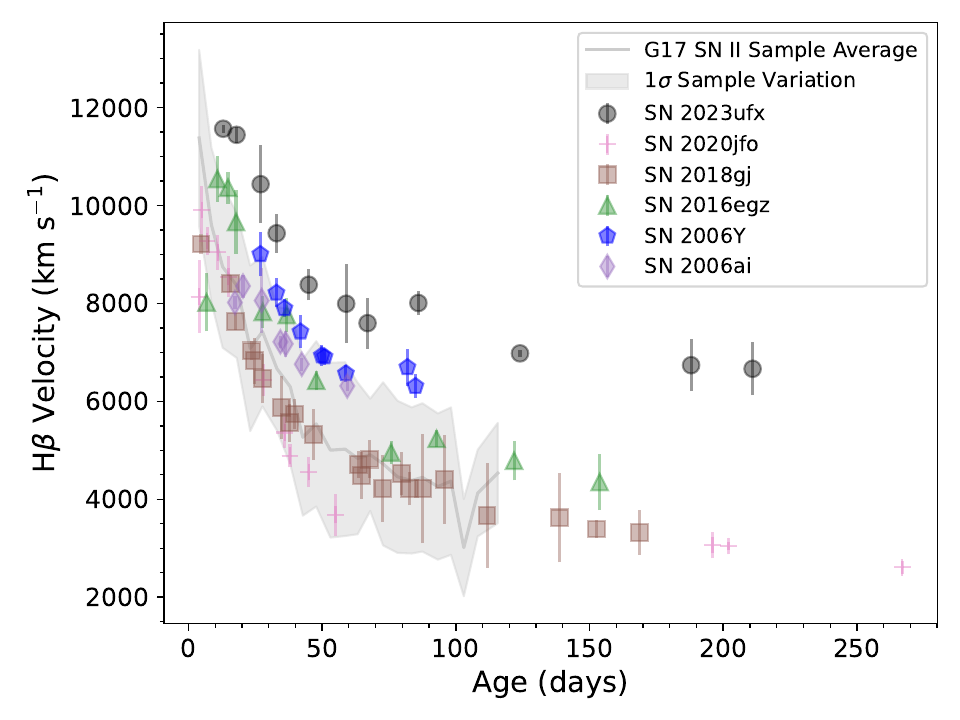}
    \caption{Comparison of H$\beta$ absorption minimum velocities with the Type II sample in \cite{Gutierrez17_first} and other SPSNe discussed in this work. The 1$\sigma$ uncertainties on the sample measurements are shaded in grey. The H$\beta$ in SN\,2023ufx is consistently at higher velocities than a typical Type II SN and all other SPSNe in literature. Reference for SPSNe data: SN\,2006Y and SN\,2006ai \citep{Gutierrez17_first}; SN\,2016egz \citep{Hiramatsu21}; SN\,2018gj \citep{Teja23}; SN\,2020jfo \citep{Teja22}.}
    \label{fig:Hbeta_velocity}
\end{figure}

\begin{figure*}
    \centering
    \includegraphics[scale=0.6] {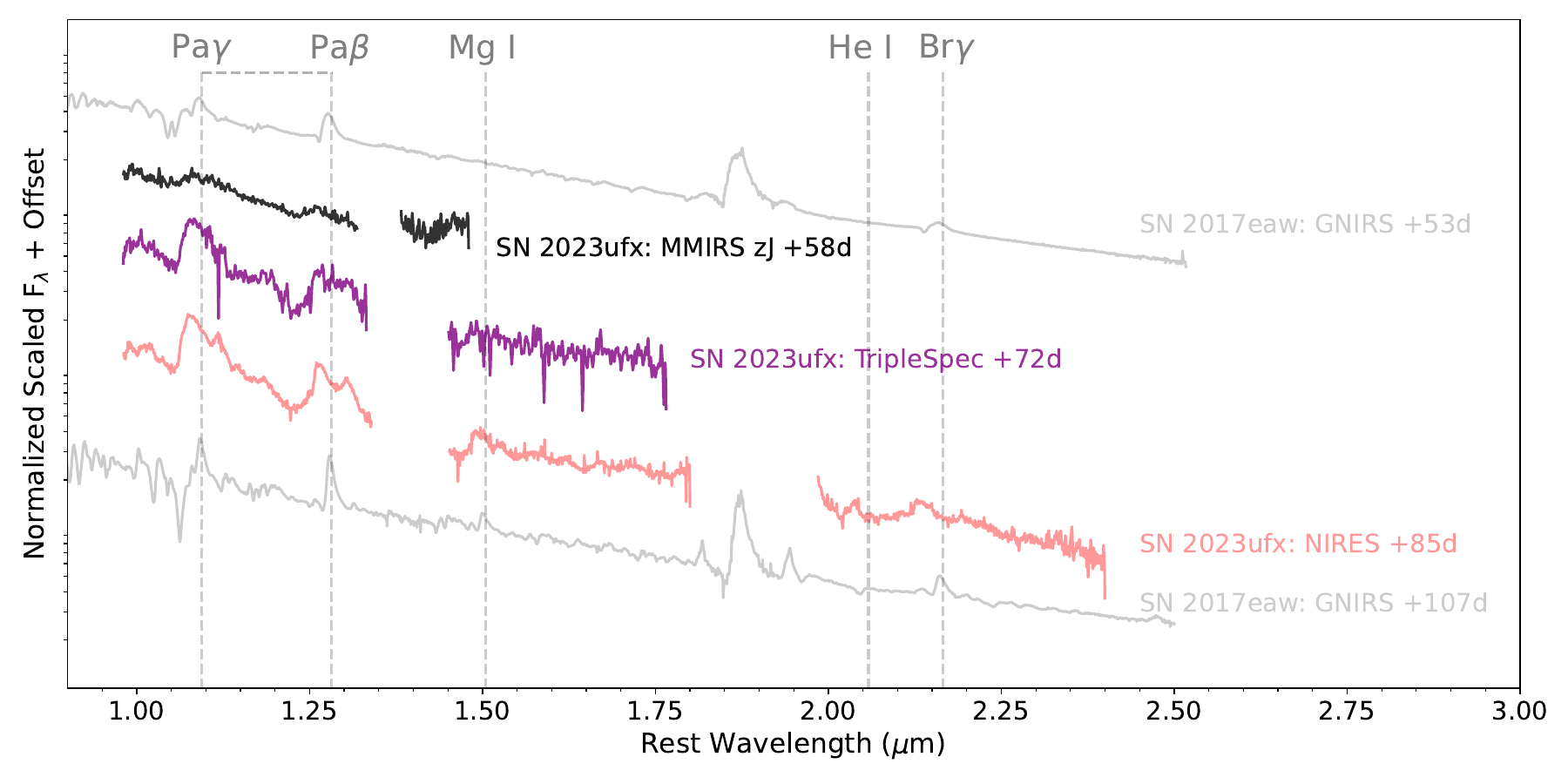}
    \caption{NIR spectra of SN\,2023ufx between 58 and 85 days. All spectra have been corrected for redshift and reddening. Regions of high telluric-absorption have been masked out. By day 81, the Pa$\gamma$ and Pa$\beta$ H lines develop a clear double peak. Two NIR spectra of the more typical SN\,IIP\,2017eaw \citep{Rho18} at closest epochs to SN\,2023ufx are plotted for comparison. }
    \label{fig:NIR_evolution}
\end{figure*}

\begin{figure}[!t]
    \centering
    \includegraphics[width=0.5\textwidth] {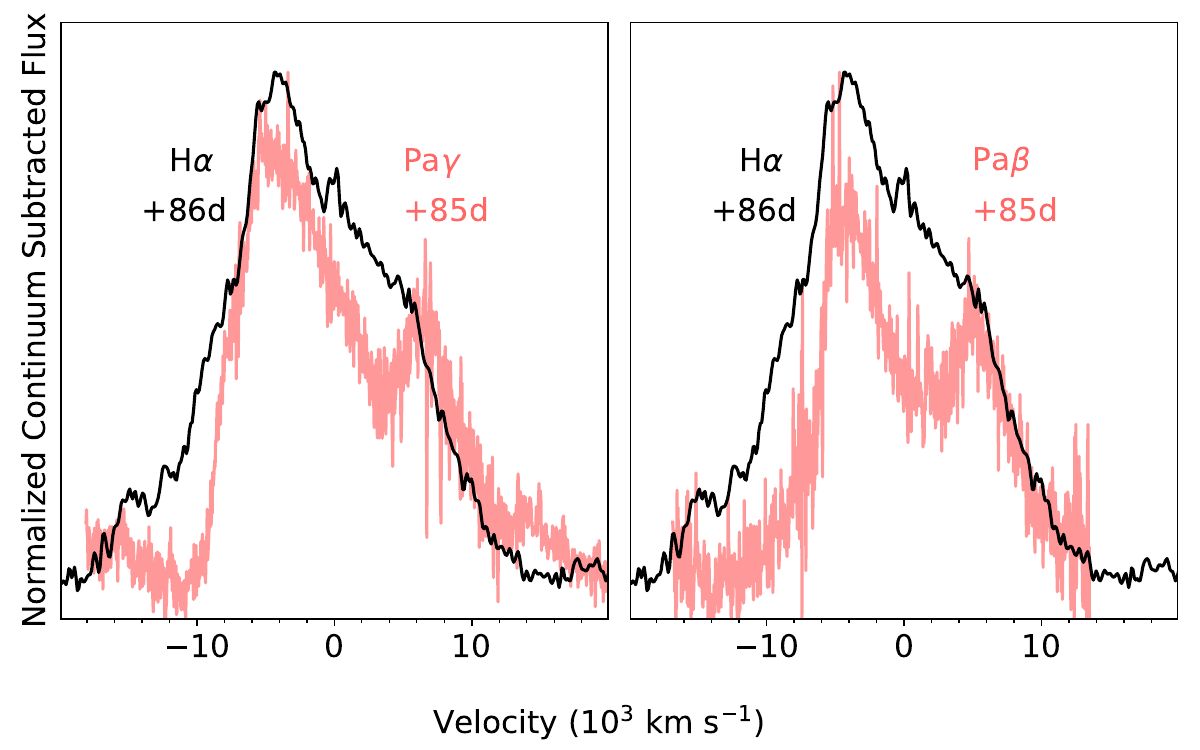}
    \caption{Comparison between optical H$\alpha$ line profile on day 86 with Pa$\gamma$ (left) and Pa$\beta$ (right) line profiles in the NIR on day 81 in the velocity space. Rest wavelengths of H$\alpha$, Pa$\gamma$, and Pa$\beta$ are assumed to be at zero velocity. Double peaks in the NIR H lines suggest that observed asymmetry in the optical could be due to the asymmetry of the H-ejecta distribution from the SN explosion. The narrow line atop the broad H$\alpha$ is possibly from the host.}
    \label{fig:H-alpha comparison}
\end{figure}

\section{Spectroscopic Evolution} \label{sec:5}
\subsection{Optical Spectra} \label{sec:5.1}
The optical spectroscopic evolution of SN\,2023ufx between days 7 and 86 is shown in Figure \ref{fig:photospheric_spectra}. The late-time evolution between days 114 and 211 are shown in Figure \ref{fig:nebular_spectra}.

At early times ($\sim$7--11 days), the optical spectra of SN\,2023ufx show a blackbody-like continuum with no emission or absorption features. We present a comparison between the available early spectra of several short-plateau SNe discussed in this work in Figure \ref{fig:optical_spectra_comparison}a. The lower luminosity SPSNe (SN\,2020jfo and SN\,2018gj) start showing the emergence of hydrogen Balmer lines at this time while the more luminous SN\,2016egz shows a similar continuum as SN\,2023ufx. As the earliest available spectrum in our study of SN\,2023ufx is at day 7, we do not observe the fleeting narrow high-ionization CSM emission lines (e.g., from H, He I, He II; also called flash features) observed in the early spectra of some SNe\,II \citep[e.g.,][]{GalYam14, smith15, Khazov16, Yaron17, Tartaglia21, Bruch21, Hiramatsu2021NatAs...5..903H, Terreran22, Bostroem23, Bruch23, Jacobson-Galan23, Hiramatsu23, Jacobson-Galan24, Dessart23, Shrestha24a}. This is strengthened by the observations of \cite{Tucker24}, where an even earlier spectrum of SN\,2023ufx at $\sim$3.5 days does not show any flash features.

As SN\,2023ufx enters the plateau phase around day $\sim$20, the spectra start becoming redder, showing hydrogen Balmer-series features, with a broad H$\alpha$ detection. The H$\alpha$ profile has a strong emission feature with negligible absorption. Among the SNe\,II samples of \cite{Anderson14} and \cite{Gutierrez14, Gutierrez17_first, Gutierrez17_second}, SPSNe 2006ai and 2006Y were noted to have the lowest absorption-to-emission strength ($a/e$) ratios. These studies found significant correlations between $a/e$ and plateau length with more IIL-like SNe having weaker ratios. Additionally, SN\,2016egz (another luminous short-plateau) was also shown to have low $a/e$ \citep{Hiramatsu21}. We show the comparison of H$\alpha$ profile with our SPSN sample during the plateau phase in Figure \ref{fig:optical_spectra_comparison}b. SN\,2023ufx has among the smallest absorption to emission strength in this sample.  For the IIL sample, several potential scenarios have been proposed to explain a low $a/e$, including significantly lower M$_\mathrm{H_\mathrm{env}}$ than typical SNe\,II \citep[e.g.,][]{Schlegel96, Gutierrez14, Gutierrez17_second}. Thus, the low $a/e$ in SN\,2023ufx is consistent with other luminous SPSNe and could be indicative of a small hydrogen envelope mass in the aftermath of partial stripping. Additionally, the H$\alpha$ profile in SN\,2023ufx is significantly broader and starts showing signs of multiple components by day $\sim$ 59 days (Figure \ref{fig:optical_spectra_comparison}b). 

One evident difference (from a typical SN II) in the optical spectral evolution of SN\,2023ufx is the absence of strong metallic absorption features in the photospheric phases (Figure \ref{fig:photospheric_spectra}). The weak absorption of Fe~II 5169 \AA\ only shows up at $\sim$59 days, while Fe~II absorption lines are evident in the early plateau phase for majority of the SPSNe (Figure \ref{fig:photospheric_spectra}). Weak metal lines could be due to low metallicity progenitor of SN\,2023ufx as suggested by \cite{Tucker24}. By comparing the optical spectra of SN\,2023ufx with the metallicity-dependent model spectra of \cite{Dessart13}, they found that the Fe II and Ca II features in SN\,2023ufx are weaker than $\sim$0.1 $Z_{\odot}$ metallicity models at comparable epochs. 

Hydrogen Balmer lines can be used to probe the expansion velocities associated with SNe\,II \citep[e.g.,][]{Gutierrez17_first}.
The evolution of the line profile of H$\alpha$ in SN\,2023ufx is significantly more rapid than other SPSNe. As the SN ends the plateau phase, the single broad H$\alpha$ peak develops two distinct emission components between days $\sim$51 (Figure \ref{fig:photospheric_spectra}) and $\sim$211 (Figure \ref{fig:nebular_spectra}). In velocity space, these emission components are at $\sim$4800 km s$^{-1}$ from the reference rest-wavelength of H$\alpha$ at 6563 \AA\ (Figure \ref{fig:optical_velocity}). 

The other Balmer lines (H$\beta$, H$\gamma$, and H$\delta$) in SN\,2023ufx show strong and  persistent absorption lines. The velocity evolution of H$\beta$ based on absorption minimum, compared with the SNe\,II sample average of \cite{Gutierrez17_first} and other SPSNe discussed in this work is shown in Figure \ref{fig:Hbeta_velocity}. Systematically the velocities in SN\,2023ufx are higher than a typical SN II at comparable epochs. While all other SPSNe fall within the 1-sigma variation of typical SNe\,II, SN\,2023ufx has significantly faster H$\beta$ throughout its evolution. This could suggest faster expansion velocities and higher explosion energies than typical SNe\,II. 

We favor comparisons of velocities with H$\beta$ absorption minimum as an insignificant absorption and the presence of multiple components in the H$\alpha$ profile introduce significant uncertainties in its velocity measurements. Differences in evolution of the other Balmer lines (mostly in absorption) compared to H$\alpha$ (mostly in emission) could be due to differences in optical depths and blending from Fe-group lines \citep{Gutierrez17_first}

In the nebular phase, the single broad H$\alpha$ peak develops into two distinct emission components between days $\sim$59 (Figure \ref{fig:photospheric_spectra}) and $\sim$211 (Figure \ref{fig:nebular_spectra}). In the velocity space, the two components are roughly at -4800 and 4800 km s$^{-1}$ with respect to the rest wavelength of H$\alpha$ (Figure \ref{fig:optical_velocity}). Over time, the profile appears to become more ``boxy' or ``flat-topped'. In addition to H$\alpha$, strong evolution is seen in [O~I]~$\lambda\lambda$6300,\,6364 and [Ca II]~$\lambda\lambda$7321,\,7339. Between days $\sim$114 and $\sim$211, the [O~I]~$\lambda\lambda$6300,\,6364 doublet strengthens dramatically, developing three distinct peaks by the latest nebular spectrum in our campaign (Figure \ref{fig:optical_velocity}). While the central peak aligns with the rest wavelength of 6300 \AA\, the two peaks on either sides are at velocities of -2800 and 2800 km s$^{-1}$ with respect to 6300 \AA. For the [Ca~II] doublet, a blue-shifted and red-shifted peak are seen at -2900 and +6000 km s$^{-1}$ with respect to the rest wavelength of 7291 \AA\ (Figure \ref{fig:optical_velocity}). The development of a flat-top could be indicative of CSM interactions powering the emission lines and nebular phase luminosity \citep[e.g., like in SN\,1993J;][]{Matheson00, Matheson00b}, while the differences in velocities between H, O, and Ca could suggest a spatially separated origin for their emission. We discuss some of the different physical geometries suggested by these nebular line profiles of H, O, and Ca in Section \ref{sec:7}.

\subsection{Near-Infrared Spectra} \label{sec:5.2}
To understand the multi-peaked H$\alpha$ profile, we can compare the optical emission features with those in NIR that are generally better isolated (e.g., Pa$\gamma$, Pa$\beta$). The NIR spectral coverage of SN\,2023ufx is shown in Figure \ref{fig:NIR_evolution}. Between day 51 and 86 when the optical H$\alpha$ starts developing a two-component profile, we observed SN\,2023ufx three times at NIR wavelengths. While the S/N was not ideal in the first two epochs, the Keck NIRES spectrum at day 85 clearly shows double peaks for Pa$\gamma$ and Pa$\beta$ line profiles (Figure \ref{fig:NIR_evolution}). We also compare the NIR spectra of SN\,2023ufx with those of the well-studied SN\,2017eaw \citep{Rho18} at nearby epochs of 53 and 107 days. The most stark difference is the presence of double-peak Paschen line profiles in SN\,2023ufx compared to the more typical single peak Paschen profiles in SN\,2017eaw (Figure \ref{fig:NIR_evolution}).

We compare the scaled and normalized Pa$\gamma$ and Pa$\beta$ profiles with H$\alpha$ in the velocity space in Figure \ref{fig:H-alpha comparison}. The similarities in the overall line profile shapes and line widths ($\sim$ 10,000 km s$^{-1}$) between them in SN\,2023ufx suggest that the developing second peak in H$\alpha$ (which is later clearly seen in nebular spectra; Figure \ref{fig:nebular_spectra}) is likely due to hydrogen emission and not blending of other species. The marginal excess on the blue side of the H$\alpha$ over Pa$\gamma$ and Pa$\beta$ could be due to the emerging presence of [O~I]~$\lambda\lambda$6300,\,6364 around day $\sim$86, which strengthens significantly in the nebular spectra (Figure \ref{fig:nebular_spectra}). While the overall line widths are similar between optical and NIR, the Paschen lines show the presence of two distinct peaks that appear to be more blended in the optical H$\alpha$ profile (Figure \ref{fig:H-alpha comparison}). This could be due to hydrogen spectral lines in the Paschen and Balmer series having transitions starting at different energy levels (n = 3 vs n = 2 in the orbitals of a Hydrogen atom) and thus potentially a different transparency to the ejecta.


\begin{figure*}
    \centering
    \includegraphics[width=\textwidth] {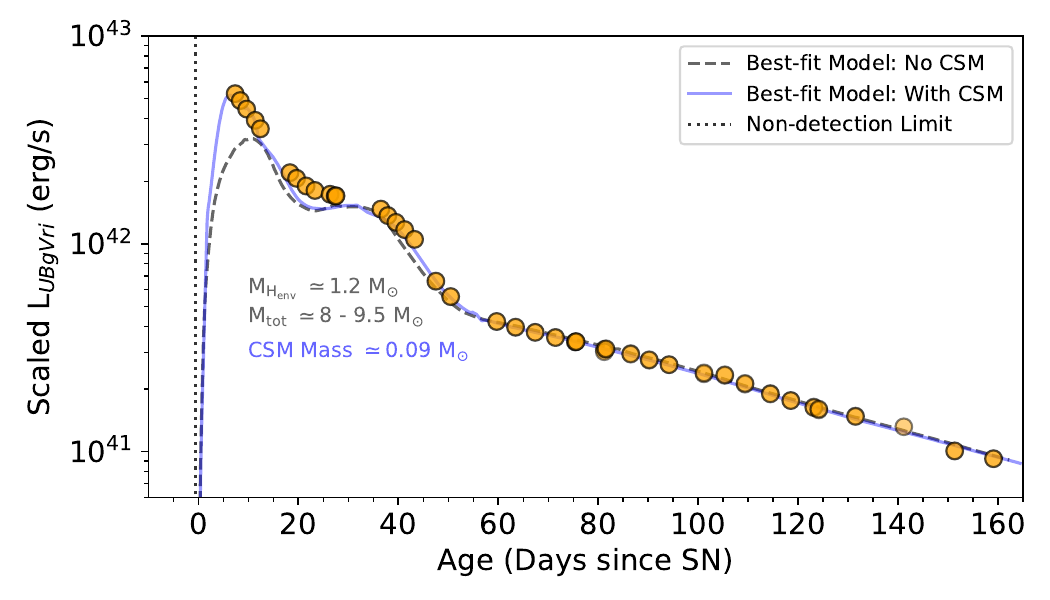}
    \caption{The pseudo-bolometric light curve of SN\,2023ufx scaled to M$_\mathrm{Ni}$ = 0.1 M$_{\odot}$ and overlaid with the ``best-fit'' (i.e., with the highest likelihood) MESA+STELLA light curves from the model grids set at M$_\mathrm{Ni}$ = 0.1 M$_{\odot}$ (see Section \ref{sec:6.1}) with and without the presence of CSM. The most likely value of the hydrogen envelope is around $\sim$1.2 M$_{\odot}$ and the total mass of the progenitor at core-collapse is $\sim$8 -- 9.5 M$_{\odot}$. CSM is required to account for the luminosity at early times ($\lesssim$10 days). The best-fit CSM model has a mass-loss rate of $\sim$3 $\times$ 10$^{-3}$ M$_{\odot}$ yr$^{-1}$ for a period of 30 yr before core-collapse. This accounts to a total of 0.09 M$_{\odot}$ CSM with a radial extent of $\sim$9.5 $\times$ 10$^{15}$ cm, associated with the rapid early decline observed in SN\,2023ufx.}
    \label{fig:MESA_STELLA_best_fit}
\end{figure*}

\begin{figure*}
    \centering
    \includegraphics[width=\textwidth] {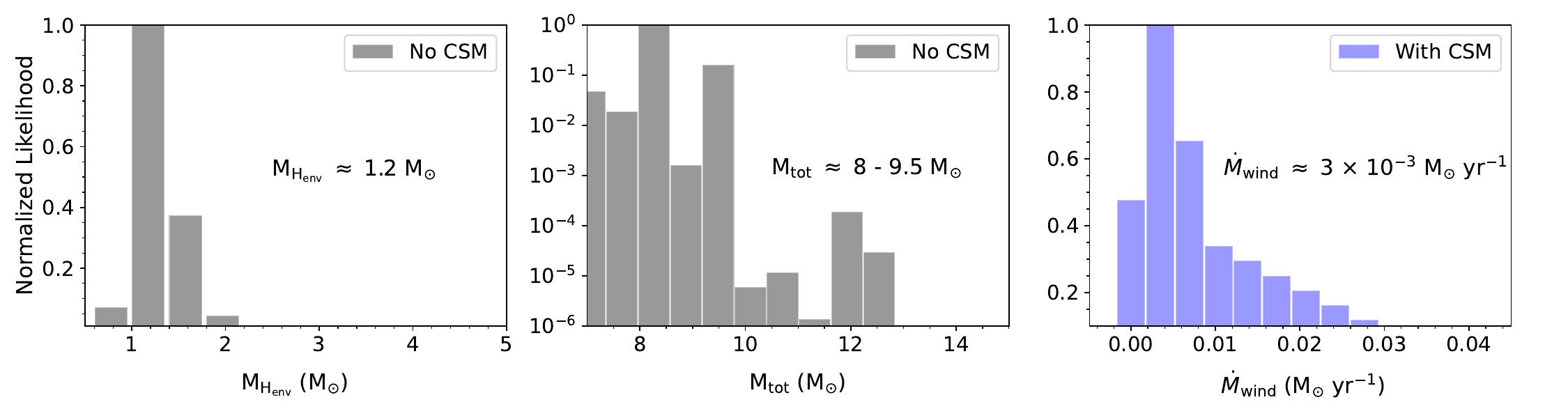}
    \caption{\textit{(Left)}: Marginal normalized likelihood distribution of M$_\mathrm{H_\mathrm{env}}$ over the full parameter space of the MESA+STELLA model grid without any CSM (see Section \ref{sec:6.1}). The distribution peaks around $\sim$1.2 M$_{\odot}$ suggesting partial stripping of the H-envelope in SN\,2023ufx. (\textit{Middle}): Marginal normalized likelihood distribution of M$_\mathrm{tot}$ over the full parameter space of the MESA+STELLA model grid without any CSM. The total mass distribution of the progenitor at core-collapse has the highest likelihood between 8 -- 9.5 M$_{\odot}$.  \textit{Right:} Marginal normalized likelihood distribution of the mass loss rate over the parameter space of the MESA+STELLA model grid where additional powering was introduced through CSM (see Section \ref{sec:6.1}). The distribution peaks around a mass loss rate of 3 $\times$ 10$^{-3}$ M$_{\odot}$ yr$^{-1}$ to explain the early decline in SN\,2023ufx. }
    \label{fig:Max_likelihood}
\end{figure*} 

\section{Progenitor Properties} \label{sec:6}

\subsection{Hydrodynamic Models: MESA+STELLA Model Grid} \label{sec:6.1}
To explore the effects of the hydrogen-rich envelope (H-envelope) being stripped during stellar evolution of SNe\,II progenitors, \cite{Hiramatsu21} constructed a large \verb|MESA| \citep{Paxton11, Paxton13, Paxton15, Paxton18, Paxton19} +\verb|STELLA| \citep{Blinnikov98, Blinnikov00, Blinnikov_Sorokina04, Blinnikov06, Baklanov05} model grid assuming a single progenitor channel (i.e., mass loss due to stellar winds alone). In the construction of the \verb|MESA|+\verb|STELLA| model grid, \cite{Hiramatsu21} consider SPSNe as a transitional class between SNe\,IIL and IIb, having a narrow range of H-envelope mass left at core-collapse (0.91 M$_{\odot}$ $\lesssim$ M$_\mathrm{H_\mathrm{env}}$ $\lesssim$ 2.12 M$_{\odot}$) with a higher than average nickel nucleosynthesis. A detailed description of the model grid and dependent parameters is presented in the Appendix of \cite{Hiramatsu21}. 

The H-envelope in the model grid is stripped arbitrarily by varying $\eta_\mathrm{wind}$ and, there exists no one-to-one mapping between the total mass left at core-collapse, M$_\mathrm{tot}$ and M$_\mathrm{ZAMS}$. Thus, following \cite{Hiramatsu21}, we split the mass at core-collapse in the models into M$_\mathrm{He_\mathrm{core}}$ and M$_\mathrm{H_\mathrm{env}}$, such that M$_\mathrm{He_\mathrm{core}}$ = M$_\mathrm{tot}$ - M$_\mathrm{H_\mathrm{env}}$ at core-collapse. The M$_\mathrm{He_\mathrm{core}}$ is less sensitive to H-rich envelope stripping and metallicity for massive stars with M$_\mathrm{ZAMS} \lesssim30 M_{\odot}$ \citep{Woosley_Weaver95, Woosley02}, although binary interaction which is not accounted for in these models, may alter the relation \citep[e.g.,][]{Zapartas19, Zapartas21}. We can then translate the estimated M$_\mathrm{He_\mathrm{core}}$ to M$_\mathrm{ZAMS}$ by using the relation between these parameters from the stellar evolution model grid of \cite{Sukhbold16}.

The model grid constructed with M$_\mathrm{Ni}$ fixed at 0.1 M$_{\odot}$ is the closest to the observed $^{56}$Ni mass of M$_\mathrm{Ni}$ $\sim$0.14 M$_{\odot}$. In order to avoid any systematic effects in the comparison between data and the model grid, we scale our pseudo-bolometric light curve to match the fixed 0.1 M$_{\odot}$ $^{56}$Ni mass of the model grid. Theoretical models have suggested that the extra heating from additional $^{56}$Ni does not generally affect the luminosity of the IIP light curve during the plateau phase \citep[e.g.,][]{Kasen_Woosley09, Bersten13, Goldberg19}. Thus, we scale the pseudo-bolometric ($UBgVri$) light curve of SN\,2023ufx constructed as described in Section \ref{sec:4.3} after the end of the plateau phase (i.e., after $t_\mathrm{PT}$) to compare with the model grid of \cite{Hiramatsu21} with the M$_\mathrm{Ni}$ fixed at 0.1 M$_{\odot}$. 

We perform $\chi^{2}$ fitting between the scaled pseudo-bolometric light curve and the complete set of model grid light curves (M$_\mathrm{Ni}$ = 0.1 M$_{\odot}$). The best-fit model has the highest likelihood value, assuming the standard definition of $\chi^{2}$ as our likelihood function. In Figure \ref{fig:MESA_STELLA_best_fit}, we show the best-fit model with the highest likelihood to the scaled pseudo-bolometric light curve. We find that the extreme short-plateau of SN\,2023ufx can be approximately reproduced with $M_\mathrm{H_\mathrm{env}}$ $\simeq$1.2 M$_{\odot}$ and a M$_\mathrm{tot}$ $\simeq$8 -- 9.5 M$_{\odot}$. We present univariate marginal likelihood distributions of M$_\mathrm{H_\mathrm{env}}$ and M$_\mathrm{tot}$ in Figure \ref{fig:Max_likelihood}. 

While the model grid depends on several other free parameters with correlations between them \citep[see][for a more detailed discussion]{Hiramatsu21}, M$_\mathrm{H_\mathrm{env}}$ and M$_\mathrm{tot}$ are our primary interest to estimate the hydrogen envelope and the He-core mass at the time of core-collapse. Based on these ``best-fit'' values, by definition the M$_\mathrm{He_\mathrm{core}}$ for SN\,2023ufx is between $\simeq$5.8 -- 8.3 M$_{\odot}$. Using the M$_\mathrm{He_\mathrm{core}}$ to M$_\mathrm{ZAMS}$ relation in \cite{Sukhbold16}, we translate this He-core mass range to M$_\mathrm{ZAMS}$ $\simeq$19 -- 25 M$_{\odot}$. These massive progenitor estimates are generally consistent with the mass estimates (M$_\mathrm{ZAMS}$ $\simeq$18 -- 22 M$_{\odot}$)  of other luminous SPSNe \citep[like SNe 2006Y, 2006ai, 2016egz;][]{Hiramatsu21}. A more massive progenitor than typical SNe\,II with a small hydrogen envelope at core-collapse suggests that significant stripping could have occurred in the evolution of the progenitor of SN\,2023ufx, increasing support for luminous SPSNe having properties that lie between SNe\,II and SNe\,IIb. One additional caveat to note is that since the plateau length of these SPSNe models can decrease when we scale the $^{56}$Ni down \citep[see Figure 6 of][]{Hiramatsu21}, our estimated hydrogen-envelope mass of $\simeq$1.2 M$_{\odot}$ is likely an upper limit, suggesting potentially an even more stripped scenario in reality.

Theoretical models have suggested that the extra heating from additional $^{56}$Ni can extend the duration of the plateau phase \citep[e.g.,][]{Goldberg19}. However, in SN\,2023ufx, it is not the lack of heating from $^{56}$Ni which is causing the shortening of the plateau as it has among the highest amount of $^{56}$Ni synthesized in SNe\,II (see Section \ref{sec:4.3}). Rather, it is likely the small H-envelope ($\simeq$1.2 M$_{\odot}$) which is responsible, though there can be degeneracies due to the influence of explosion energy and progenitor radius. It has also been noted through theoretical models that r-process enrichment in H-rich supernovae could also truncate the plateau duration length as the H-rich photosphere enters the r-process-enriched layers \citep{Patel24}. In such a scenario it is expected that the late-time ($\sim$200 days) luminosity would be in excess of that expected from radioactive decay chain, due to excess heat trapped by the high-opacity r-process-enriched core. However, for SN\,2023ufx the luminosity continues to decline faster than radioactive decay chain (Figure \ref{fig:SPSNe_comparison}) suggesting that the shortening of the plateau may not be due to r-process enrichment. Thus, a combination of high $^{56}$Ni mass and a short-plateau duration can rule out alternative theories for the observed trends in SN\,2023ufx. As noted before, a high $^{56}$Ni mass could be overestimated due to luminosity contributions from CSM interaction at late times.

While the plateau and radioactive phase of the observed light curve can be explained with these models, it is clear from the comparisons that the first $\sim$10 days of the pseudo-bolometric LC is significantly underestimated in the best-fit model (Figure \ref{fig:MESA_STELLA_best_fit}). Hydrodynamic modeling of SNe\,II have suggested that the early phases are dominated by the presence of dense CSM surrounding the progenitor star \citep[e.g.,][]{Morozova18}. Thus, we propose CSM interaction as a possible source for the early excess observed in SN\,2023ufx. As before, we compare with a light curve model grid from \cite{Hiramatsu21} where a wind density profile $\rho_\mathrm{wind} = \dot{M}_\mathrm{wind}/4 \pi r^{2}\nu_\mathrm{wind}$ is added to a subset of the explosion models resulting in SPSNe (i.e., for models with 0.91 M$_{\odot}$ $\lesssim$ M$_\mathrm{H_\mathrm{env}}$ $\lesssim$ 2.12 M$_{\odot}$). With a H-envelope of $\sim$1.2 M$_{\odot}$ in SN\,2023ufx, these models are valid for our case.  Here, $\dot{M}_\mathrm{wind}$ is a constant wind mass-loss rate and $\nu_\mathrm{wind}$ is the wind velocity for an arbitrary amount of time, $t_\mathrm{wind}$ (i.e., $\dot{M}_\mathrm{wind}$ $\times$ $t_\mathrm{wind}$ = M$_\mathrm{CSM}$). The model grid includes $\dot{M}_\mathrm{wind}$ values between 10$^{-5}$ and 10$^{-1}$ M$_{\odot}$ yr$^{-1}$ and two $t_\mathrm{wind}$ values at 10 and 30 yr for each short-plateau SN model, assuming a typical $\nu_\mathrm{wind}$ = 10 km s$^{-1}$ \citep[e.g.,][]{Moriya11}. Note that these mass-loss rates are much higher than observed mass-loss rates of massive stars \citep{smith14,Beasor20}.

Similar to the comparisons with the non-CSM model grid, we perform $\chi^{2}$ fitting between the observed pseudo-bolometric light curve and the CSM model grid. The best-fit light curve with additional powering from CSM is shown in Figure \ref{fig:MESA_STELLA_best_fit}. We find that early excess in the observed pseudo-bolometric light curve can be explained with $\dot{M}_\mathrm{wind}$ $\sim$ 3 $\times$ 10$^{-3}$ M$_{\odot}$ yr$^{-1}$ for a period of $t_\mathrm{wind}$ = 30 yr. As the best-fit models show, the additional power from the CSM in these models affects only the early evolution before the plateau phase. So, the estimated M$_\mathrm{H_\mathrm{env}}$ and M$_\mathrm{tot}$ from the best-fit models with and without CSM will be identical.  In Figure \ref{fig:Max_likelihood}, we show the univariate marginal likelihood distribution of $\dot{M}_\mathrm{wind}$. The best-fit $\dot{M}_\mathrm{wind}$ $\sim$ 3 $\times$ 10$^{-3}$ M$_{\odot}$ yr$^{-1}$  suggests enhanced mass-loss \citep[a few orders of magnitude higher than the typical RSG steady-state mass loss  $\dot{M}_\mathrm{wind}$ $\sim$ 10$^{-6}$ M$_{\odot}$ yr$^{-1}$;][]{Beasor20} for a few decades before the SN explosion. This corresponds to a total CSM mass of M$_\mathrm{CSM}$ $\sim$ 0.09 M$_{\odot}$ for a period of 30 yr. For the assumed CSM velocity ($\sim$10 km s$^{-1}$) and the best-fit CSM duration, $t_\mathrm{wind}$ = 30 yr, the radial extent of this CSM is $\sim$9.5 $\times$ 10$^{15}$ cm. As a sanity check, for a typical SN shock velocity of $\sim$10,000 km s$^{-1}$, this distance can be swept in $\sim$10 days, which is consistent with the observed rapid decline of the light curve for the first $\sim$10 days (Figures \ref{fig:photometry}, \ref{fig:MESA_STELLA_best_fit}).

Dense CSM around massive stars through mass-loss is expected to produce narrow high-ionization emission lines in the early spectra \citep{Smith17,Dessart17}. As presented in Section \ref{sec:5.1}, early spectra of SN\,2023ufx do not show such features despite an enhanced mass-loss rate of $\sim$ 3 $\times$ 10$^{-3}$ M$_{\odot}$ yr$^{-1}$ suggested from light curve modeling. One reason why no corresponding flash features were observed could be simply a lack of early enough spectrum (earliest spectrum in our work is at day 7 and day 3.5 in \cite{Tucker24}). Another possibility is that the CSM responsible for a rapid decline in the observed light curve is asymmetrically distributed around the progenitor where ejecta CSM interactions still happen but narrow lines are not observed from several viewing angles \citep[see][]{Smith17}.

The progenitor ZAMS mass inferred from hydrodynamic modeling ($\simeq$19 -- 25 M$_{\odot}$) is higher than typical RSG progenitors (ZAMS mass $\simeq$8 -- 17 M$_{\odot}$) identified to be associated with SNe\,II through pre-explosion imaging \citep[][]{Van_Dyk03, Smartt09, Smartt15, Van_Dyk17}. The high progenitor mass inferred for SNe 2006Y, 2006ai, and 2016egz highlighted a possibility that partially-stripped massive progenitors end their evolution as RSGs and explode as SPSNe, rather than getting stripped enough to be SESNe or collapse to form black holes \citep{Hiramatsu21}. The even higher progenitor mass scenario coupled with the shortest-plateau among IIP, suggests that SN\,2023ufx could be consistent with this physical picture. At the same time, much lower progenitor masses were inferred for SPSNe 2020jfo \citep[ZAMS mass $\sim$12 M$_{\odot}$;][]{Teja22} and 2018gj \citep[ZAMS mass $\leq$13 M$_{\odot}$;][]{Teja23}. The clear distinguishing property among these two classes of SPSNe could be their luminosities (See Figure \ref{fig:SPSNe_comparison}), suggesting the possibility of two distinct progenitor channels to form a short-plateau in the observed light curve. Thus, studying a larger sample of SPSNe and constraining their progenitor masses could provide some insights into the proposed RSG problem \citep{Smartt15} where there is an apparent discrepancy between progenitor masses of SNe\,II and our understanding of RSG evolution.

Another caveat to note is that if RSG mass loss is metallicity dependent \citep[like that of O and B stars; e.g.,][]{Vink00, Vink01, Mokiem07, Vink21}, some interacting binary progenitor systems could also be a plausible scenario for formation of a short plateau in SNe\,IIP \citep{Eldridge17, Eldridge18}. Most SPSNe including SN\,2023ufx have a sub-solar host metallicity \citep[][; also see Section \ref{sec:2}]{Hiramatsu21, Tucker24}. However, recent observational and theoretical studies of RSGs have also suggested their wind mass-loss rates may be independent of metallicity \citep[e.g.,][]{Goldman17, Chun18, Gutierrez18, Yang23, Antoniadis24}. Thus it is possible that SN\,2023ufx with its short plateau has formed from massive single-star evolution. This is consistent with the physical picture discussed for other luminous SPSNe 2006Y, 2006ai, and 2016egz in \cite{Hiramatsu21}. A massive progenitor (ZAMS mass $\simeq$8--17 M$_{\odot}$) was suggested for the SN\,IIP\,2015bs with the lowest metallicity of any SNe\,II \citep{Anderson18}. These results suggest that a larger statistical investigation of both RSG and SN II populations at various metallicities is required to robustly distinguish between the formation channels of SPSNe.



\subsection{Progenitor Mass: Nebular Spectroscopy} \label{sec:6.2}

\begin{figure}
    \centering
    \begin{tabular}{c}
     \includegraphics[width=0.5\textwidth] {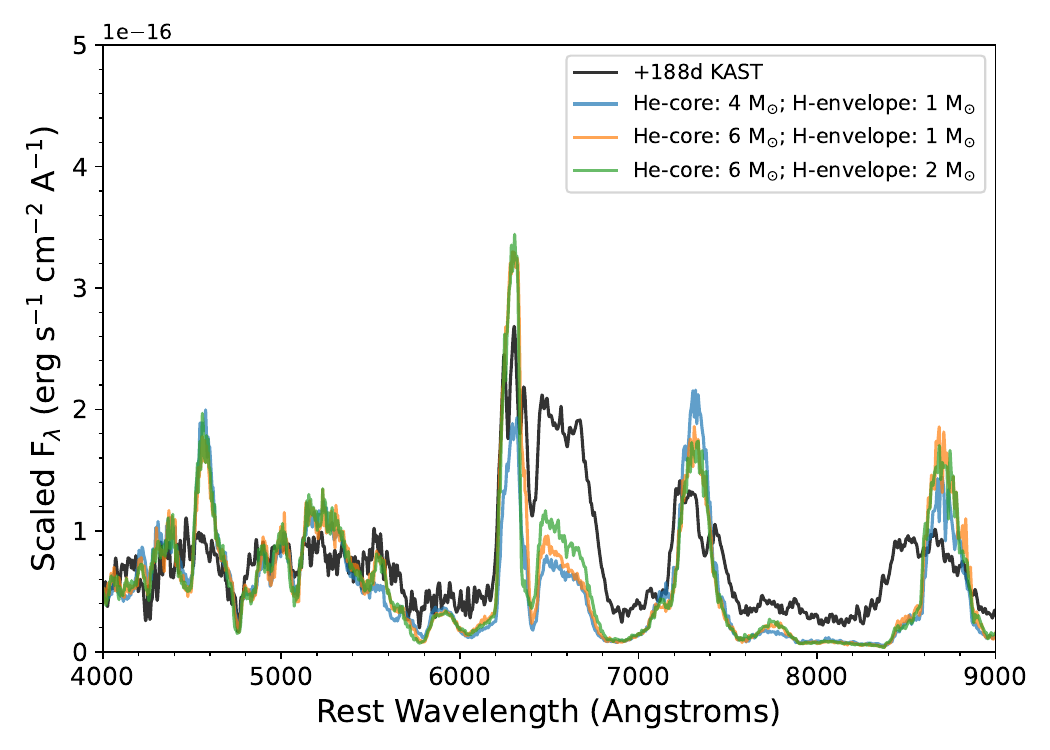}   \\
     \includegraphics[width=0.5\textwidth] {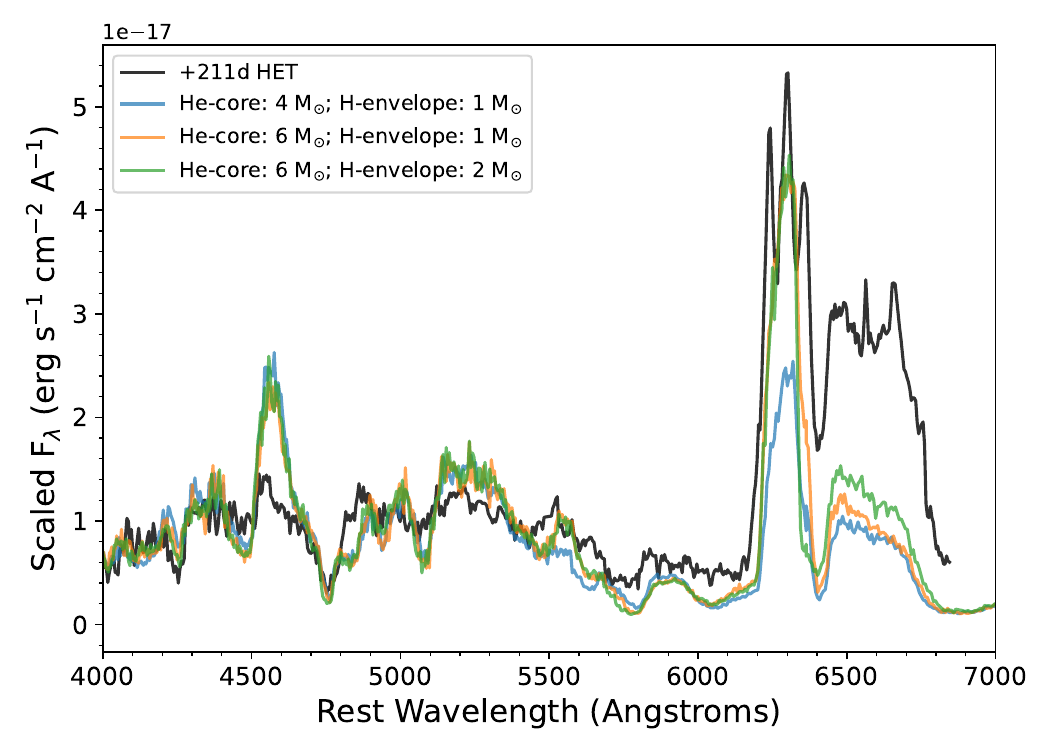} 
    \end{tabular}
    \caption{Nebular spectra at days 188 (top) and 211 (bottom) compared with three synthetic Ib models at day 200, assuming a He-core mass of 4 M$_{\odot}$ with an additional hydrogen envelope mass of 1 M$_{\odot}$, and 6 M$_{\odot}$ He-core with additional hydrogen envelope masses of 1 and 2 M$_{\odot}$, respectively. The observed strength of the [O~I] doublet is most similar to the models with a He-core mass of 6 M$_{\odot}$, though none of the models can account for the observed strength of H$\alpha$.}
    \label{fig:nebular_modeling}
\end{figure}

As the SN evolves into the nebular phase, the ejecta become optically thin, revealing the inner structure of the SN. From modeling the observed spectra at nebular phases, the strength of the [O~I]~$\lambda\lambda$6300,\,6364 doublet is known to be a good indicator of the ZAMS mass of the progenitor \citep[e.g.,][]{Jerkstrand14}. 

Given SN\,2023ufx has an intermediately sized hydrogen envelope (between $\sim$1 -- 2\, M$_{\odot}$), neither the standard comparisons with theoretical nebular spectral models for Type IIP with a large hydrogen envelope \citep{Jerkstrand14} nor Type IIb spectral models with minimal hydrogen envelope \citep{Jerkstrand15} are in the right regime for constraining the progenitor mass. Therefore, we construct a Non-Local Thermodynamic Equilibrium (NLTE) spectral synthesis model following the prescription in \cite{Barmentloo24}. 

We assume that the supernova may be approximated as a spherically symmetric, neutrino driven explosion. For this assumption, three Type Ib models with initial helium core masses of M$_\mathrm{He_\mathrm{core, init}}$ = 4.0, 6.0, and 8.0 M$_{\odot}$ as used in \cite{Barmentloo24}  \citep[themselves taken from][]{Woosley19, Ertl20} were used as the initial setup. At explosion, these masses have decreased through mass-loss to 3.2, 4.4, and 5.6 M$_{\odot}$. These models are divided into seven different compositional zones in the core and envelope. To these helium cores, hydrogen envelopes of masses 1 and 2 M$_{\odot}$ were attached, with the same chemical composition as in \cite{Jerkstrand15}, based on models for SN\,1993J from \cite{Woosley94}. The choice of envelope masses is based on our independent estimates from light curve analysis (see Section \ref{sec:6.1}). Following the velocity profile of H$\alpha$ (Figure \ref{fig:optical_velocity}), this mass was distributed between 6000 and 11,000 km s$^{-1}$, following a $\rho \propto V^{-6}$ density profile. The core zones are kept at the same velocities ($\sim$4500 km s$^{-1}$) as in \cite{Barmentloo24}, which is required by our spectra to reproduce the broad observed lines from these regions (i.e. [O~I] and [Ca~II] doublets). This leaves room for the He/C- and He/N-envelope layers only between $\sim$4500 and 6000 km s$^{-1}$, making these zones denser than in \cite{Barmentloo24}. Of these, only the He/C layer is partially mixed (18, 10, and 10\% for M$_{\mathrm{He}_\mathrm{core, init}}$ = 4.0, 6.0, and 8.0 M$_{\odot}$, respectively) into the core, with the He/N- and H-layers left unmixed. This mixing prescription is adopted based on the work by \cite{Shigeyama94}, who found that for hydrogen envelope masses of 1 - 2 M$_{\odot}$, no significant in-mixing of the hydrogen layer should occur. Additionally, introducing mixing produces more parabolic line profiles that do not match the observed asymmetric line profiles (Figure \ref{fig:nebular_spectra}).
A final consequence of this setup is that the required explosion energies are higher than models without the hydrogen envelope. For example, in the 6 M$_{\odot}$ He-core model, adding hydrogen envelopes of 1 and 2 M$_{\odot}$ increases the explosion energy of the bare helium star from 1.07 B to 1.1 and 1.7 B, respectively.

We compare the two nebular spectra of SN\,2023ufx at days $\sim$188 and $\sim$211 with one M$_\mathrm{He_\mathrm{core}}$ = 4.0 M$_{\odot}$ and two M$_\mathrm{He_\mathrm{core}}$ = 6.0 M$_{\odot}$ models (with hydrogen envelope masses between 1--2 M$_{\odot}$) in Figure \ref{fig:nebular_modeling}. The data and model are empirically scaled by integrating the total flux between 4000 and 6000 \AA. We do not include the models with He-core mass of 8.0 M$_{\odot}$ as they significantly overshoot the observed [O~I] emission feature strengths. 

Interestingly, even after accounting for 1 -- 2 M$_{\odot}$ of H, none of the models can reproduce the observed strength of H$\alpha$ in the nebular phases (Figure \ref{fig:nebular_modeling}). As independent light curve modeling suggests a hydrogen envelope mass of $\simeq$1.2 M$_{\odot}$ and SN\,2023ufx has the shortest observed plateau length among all SNe\,IIP, increasing the hydrogen envelope mass further in our nebular models is not well motivated. One scenario to explain the additional H$\alpha$ luminosity could be from CSM interaction.  In SN 1993J, the ejecta interaction with CSM powered the boxy H$\alpha$ line luminosity almost exclusively at early phases, while the light curve still declined and other lines remained little affected \citep[e.g.,][]{Houck96, Matheson00, Fransson05}.  In the case of SN\,2023ufx, the overall light curve continues to decrease in luminosity at nebular phases, but H$\alpha$, [Ca~II], and Ca II triplet spectral features all develop ``boxy" signatures with time suggesting a possibly declining interaction over time scenario. Also, as we discuss in the next section, these developments (including the rapid evolution of [O~I] doublet) could also be due to an asymmetric distribution of ejecta with H-rich material being ejected the farthest out and / or an asymmetric distribution of CSM. The models also significantly overestimate the strength of Mg~I] 4571 \AA\ line profile. This discrepancy between SN\,2023ufx data and models around Mg~I] 4571 could be due to the assumptions on the specific ionization balance of Mg and also on the adopted line blocking operating in the early nebular phases, especially at blue wavelengths \citep[see][]{Jerkstrand15}.  

Despite the unique line profiles of SN\,2023ufx and the caveats associated with our models, a high initial He-core mass of at least $\sim$6 M$_{\odot}$ is needed to account for the strength of the [O~I] doublet by day 211 (Figure \ref{fig:nebular_modeling}). One caveat to note though is that there is evidence of potential CSM interaction at late-times, so the [O~I] luminosity could also be partly powered by it. The model with initial He-core mass of 4 M$_{\odot}$ under-estimates and 8 M$_{\odot}$ over-estimates the [O~I] strength, suggesting the initial He-core mass to be $\sim$6 M$_{\odot}$. This is consistent with the lower limits from independent light curve modeling analysis and suggests the progenitor of SN\,2023ufx was massive. Based on massive Helium star evolutionary models that connect He-core mass with the ZAMS mass, a $\sim$6 M$_{\odot}$ He-core can map to massive progenitor ZAMS masses of $\simeq$19 -- 25 M$_{\odot}$ \citep[see Figure 1 in][]{Woosley19}, which we independently inferred from our light curve modeling (See Section \ref{sec:6.1}). Though, the exact connection between the two can depend on when the H-envelope was lost, which is determined by whether the He core does or does not grow through shell burning.

Our initial assumptions of 1-D spherically symmetric explosion in constructing the NLTE models also might not be ideal to explain the unique line profile features in SN\,2023ufx. Nevertheless, our independent light curve modeling clearly suggests that SN\,2023ufx exploded with a small hydrogen layer, too small to be properly treated as a typical SN II and too large to be a SN Ib with additional H introduced like in our nebular spectral models. A more thorough construction of synthetic nebular models with careful consideration of hydrogen ejecta distributions and velocities to reproduce the unique line profiles of SN\,2023ufx is beyond the scope of the current paper and will be part of future work.

\section{Asymmetric Explosion} \label{sec:7}

There has been growing evidence that most CCSNe are inherently asymmetric in their explosion mechanisms \citep[e.g.,][]{Janka07, Maeda08, Lopez09, Chornock10, Wongwathanarat13, Grefenstette14, Grefenstette17, Larsson23, Milisavljevic24, vanBaal24, Shrestha24b}. The distribution of radioactive $^{44}$Ti, synthesized in the core-collapse that formed the supernova remnant Cassiopeia A was revealed to be from a highly asymmetric bipolar explosion \citep{Grefenstette14, Grefenstette17}.  

As the SN becomes more nebular and the optical thickness of the ejecta decreases, the line profile diagnostics can give insights into the ejecta geometry causing that emission. From observations of large samples of SESNe, it has been shown that nebular line profile evolution of [O~I]~$\lambda\lambda$6300,\,6364 for a large fraction of SNe Ib/c shows asphericity in the ejecta distribution \citep[e.g.,][]{Maeda08, Modjaz08, Taubenberger09, Milisavljevic10, Fang24}. 

Late-time spectra of SN\,2023ufx between days $\sim$120 and $\sim$188, show significant changes in O, H, and Ca emission lines (see Section \ref{sec:5.1}), with each line suggesting the presence of a blue-shifted and red-shifted component. Thus, one plausible explanation for these observations would be a combination of ejecta interaction with an asymmetric CSM \citep[e.g., like in SN\,1993J;][]{Matheson00, Matheson00b} and / or an asymmetry in the geometry of the explosion, propelling material preferentially in directions along our line-of-sight.
\begin{figure}
    \centering
    \includegraphics[width=0.5\textwidth] {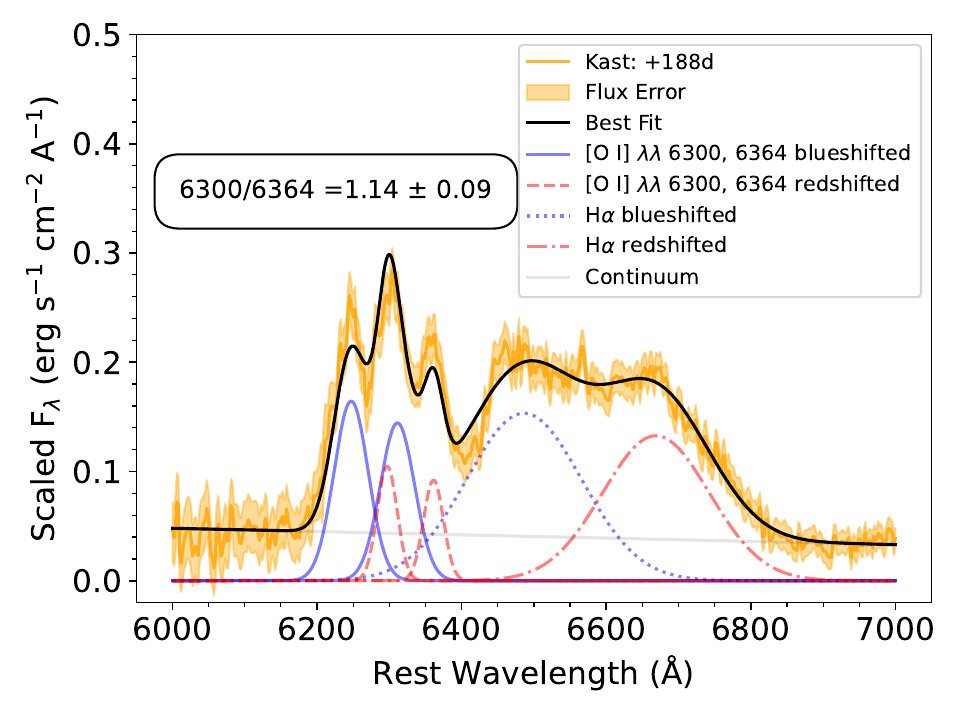}
    \caption{Composite multi-component fit of [O~I] $\lambda \lambda$ 6300, 6354 and H$\alpha$ for the nebular spectrum at day 188. The best-fit line ratio between the individual emission lines in the [O~I] doublet is 1.14 $\pm$ 0.09. The [O~I] doublet and H$\alpha$ both have a blue- and red-shifted component which could be indicative of an asymmetric distribution of ejecta in SN\,2023ufx.}
    \label{fig:OIdoublet_Halpha_d188}
\end{figure}

\begin{figure}
    \centering
    \includegraphics[width=0.5\textwidth] {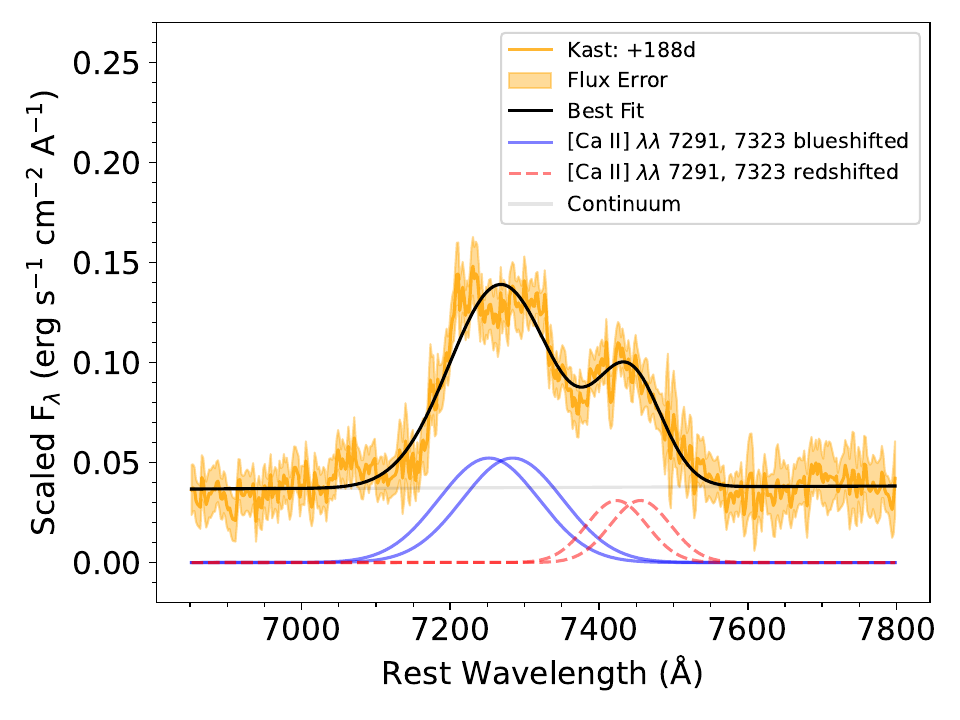}
    \caption{Two-component fit of [Ca~II] $\lambda \lambda$ 7291, 7323 for the nebular spectrum at day 188. Like O and H, the emission line profile of Ca is also fit with a blue- and red-shifted profile supporting the possibility of asymmetric explosion in SN\,2023ufx. }
    \label{fig:CaIIdoublet}
\end{figure}

To study the developing emission peaks in [O~I]~$\lambda\lambda$6300,\,6364 and H$\alpha$, we simultaneously fit both the [O~I] doublet and H$\alpha$ profiles at nebular phases. For the [O~I] doublet, we adopt the following model assumptions: 
\begin{enumerate}
    \item The observed three peaks are fit with the sum of a blue-shifted and red-shifted doublet where the Full-width at Half Maximum (FWHM) is fixed between the two lines within each doublet ($\lambda$6300 and $\lambda$6364).
    \item The distance between the central wavelengths associated with the two lines in the doublet is fixed at 64 \AA.
    \item We assume that the ratio between the $\lambda$6300 and $\lambda$6364 lines in the doublet will always be between 1:1 and 3:1 for both the blue- and red-shifted components, based on predictions and observations for SNe\,II \citep[e.g.,][]{Fransson_Chevalier89, Spyromilio91}. This implies that measuring the parameters of the $\lambda$6300 line in both the doublets suffices as the parameters of the $\lambda$6364 line are then tied.
\end{enumerate}
For H$\alpha$, we consider two Gaussian components, with their central wavelengths, FWHMs, and amplitudes as free parameters. Thus, the composite model has 12 free parameters (Amplitudes and FWHMs for blue- and red-shifted [O~I] doublets, wavelength associated with the $\lambda$6300 emission line in both [O~I] doublets; Amplitudes, FWHMs, and central wavelengths associated with the blue- and red-shifted H$\alpha$ profile). 

We find that the observed [O~I]~$\lambda\lambda$6300,\,6364 with three distinct peaks on day 188 is best-fit with a combination of blue- and red-shifted doublets (see Figure \ref{fig:OIdoublet_Halpha_d188}). Within each doublet, the ratio between the individual emission lines (6300/6364) is best fit at 1.14 $\pm$ 0.09, suggesting that at day 188, the O-ejecta is not yet optically thin \citep[e.g.,][]{Fransson_Chevalier89}. The blue- and red-shifted $\lambda$6300 lines of the [O~I] doublet have best-fit FWHM values of 865 $\pm$ 65 km s$^{-1}$ and 784 $\pm$ 60 km s$^{-1}$. The broad H$\alpha$ profile is best-fit with two Gaussians, one each accounting for the blue- and red-shifted peaks. They have an FWHM of 3627 $\pm$ 204 km s$^{-1}$ and 3373 $\pm$ 140 km s$^{-1}$, respectively. Similar velocities between blue- and red-shifted components within [O~I] and H$\alpha$ line profiles can indicate an axial symmetry in clumps moving toward and away from the observer. Additionally, H emission is possibly coming from a much faster moving ejecta than O, suggesting a spatially separated origin for the two species. 

As discussed in Section \ref{sec:6.2}, the H$\alpha$ profiles at nebular phases in SN\,2023ufx have a flat-top like feature (see Figure \ref{fig:nebular_spectra}). A flat-topped H$\alpha$ can be associated with CSM interaction in a ring-like or shell of H around the progenitor which could be powering the luminosity of H$\alpha$ \citep[e.g.,][]{Chevalier_Fransson94}. Usually in SNe\,IIP, these boxy features in H$\alpha$ associated with strong CSM interaction are observed at significantly later epochs: e.g., after $\sim$785 days in SN\,2004et \citep{Kotak09, Maguire10}, SN\,2008jb \citep{Prieto12}, SN\,2013ej \citep{Mauerhan17}, iPTF14hls \citep{Andrews_Smith18, Sollerman19}, and SN\,2017eaw \citep{Weil20}, while the corresponding phase is only $\sim$200 days for SN\,2023ufx. Also, recent spectro-polarimetric studies of hydrogen-rich SNe\,II have revealed large-scale asymmetries in their helium cores at core-collapse \citep[e.g.,][]{Nagao19, Nagao24, Shrestha24}. In the case of SN\,IIP\,2017gmr an unprecedented and extended aspherical explosion was noted through early polarization observations, suggesting that asymmetries could be present not only in the helium core but also in a substantial part of the hydrogen envelope \citep{Nagao19}. Thus, we cannot completely discriminate between the several origin scenarios and their combinations (asymmetric ejecta vs. ring-like CSM vs. asymmetric CSM) for explaining the observed H$\alpha$ emission, based solely only on our two nebular spectra. However, based on velocity stratification between H and O, and similarities in the velocities of blue- and red-shifted components within the doublet, O is likely to have been ejected preferentially along our line-of-sight.

Assuming a similar geometry for the Ca ejecta as O (in case of an line-of-sight explosion), the observed [Ca II]~$\lambda \lambda$7291,\,7323 can be fit with the combination of a blue- and red-shifted doublets (Figure \ref{fig:CaIIdoublet}) where ratios between the individual lines were fixed at 1:1. We find that the blue- and redshifted $\lambda$7291 lines in [Ca~II] doublet have FWHMs of 2772 $\pm$ 70 km s$^{-1}$ and 1660 $\pm$ 90 km s$^{-1}$. A caveat to note for [Ca II] is that the red-shifted peak / shoulder (at a shift of $\sim$6000 km s$^{-1}$ with respect to rest wavelength at $\lambda$7291; see Figure \ref{fig:optical_velocity}) could be due to the presence of [Ni~II] 7378 \AA\ \citep{Jerkstrand15}. A combination of [Ni~II] 7378 \AA\ and weak [Fe~II] 7388 \AA\ emission was reported for the short-plateau SN\,2020jfo \citep{Teja22}. This could explain the differences in the blue- and red-shifted FWHMs in [Ca~II] doublet. Introducing these lines in our line profile modeling does not improve the model fits statistically, but the lack of similarly high-velocity red-shifted emission features in O and H makes it likely that [Ca~II] line profile has contamination from [Ni~II] 7378 \AA.

In CCSNe, generally, the amount of Ca synthesized in the ejecta is insensitive to the ZAMS mass of the progenitor, while mass of O-core depends on it, making the [Ca~II] / [O~I] flux ratio especially important as an indicator of the progenitor mass \citep[e.g.,][]{Fransson_Chevalier89}. From our modeling of the nebular spectrum on day 188, we find the ratio of fluxes [Ca~II] / [O~I] = 0.93. A value of [Ca~II] / [O~I] = 1.43 was adopted as a rough division between SNe\,II and SESNe \citep[with lower values being closer to SESN with larger O-core mass;][]{Jerkstrand17, Gutierrez20, Hiramatsu21}. Thus, the low [Ca~II] / [O~I] in SN\,2023ufx suggests a high O-core mass, supporting a massive progenitor scenario. In fact, any potential contamination to the [Ca~II] line profile will only further decrease the ratio, strengthening our conclusion that the progenitor of SN\,2023ufx was more massive than a typical SN II. This is generally in agreement with the progenitor masses of SPSNe being between that of SNe\,II and SESNe \citep[see Figure 12;][]{Hiramatsu21}. A similar ``transitional'' nature between SN II and SESN based on the [Ca~II] / [O~I] ratio has been previously noted for SNe 2015bs \citep{Anderson18} and 2017ivv \citep{Gutierrez20}.  An important caveat to note is that the [Ca~II] / [O~I] ratio could be affected by the explosion energy due to more emission from synthesized Ca than the pre-existing Ca in the H-rich envelope as commonly seen in SESNe \citep{LI_McCray93, Maguire12, Jerkstrand15, Jerkstrand17}. Nevertheless, this line ratio provides a qualitative estimate of the massive progenitor of SN\,2023ufx, which is consistent with our results from independent light curve analysis as presented in Section \ref{sec:6.1}.


\begin{figure}
    \centering
    \includegraphics[width=0.5\textwidth] {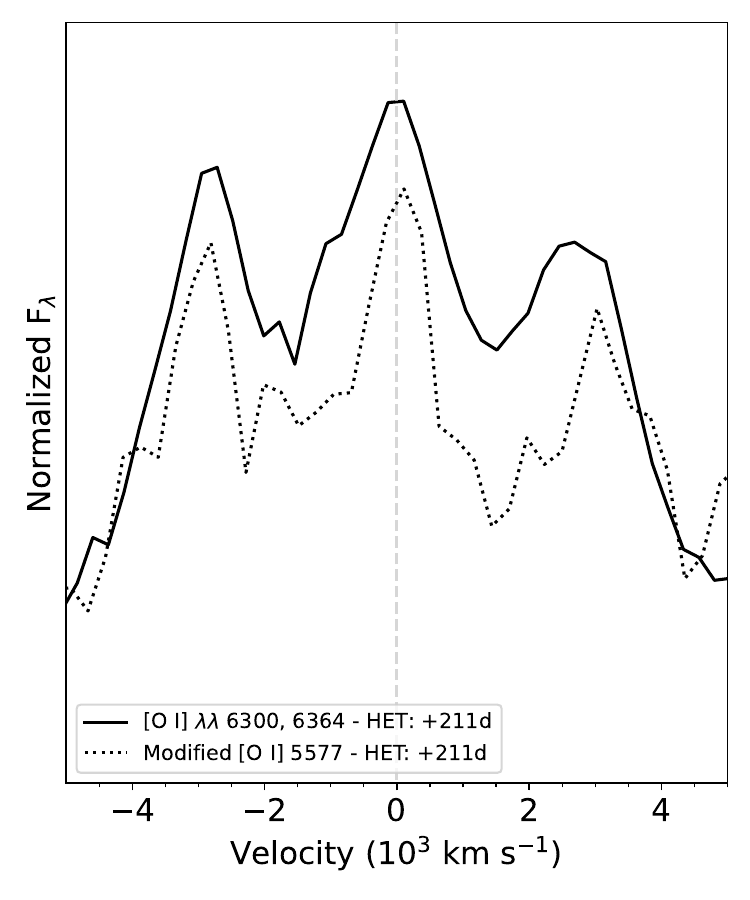}
    \caption{Comparison between [O~I] $\lambda$$\lambda$6300,6364 and modified [O~I] $\lambda$5577. Three peaks in the modified [O~I] $\lambda$5577 suggests that the observed three peaks in the [O~I] $\lambda$$\lambda$6300,6364 profile comprises mostly of oxygen, ejected preferentially toward and away from our frame of reference. A vertical line shows the zero velocity at 6300 \AA}
    \label{fig:OIDoublet and OISinglet}
\end{figure}

As [O~I] $\lambda$$\lambda$6300,6364 in the nebular phase shows multiple peaks, it is important to check for contamination from other species. To do this, we compare the [O~I] doublet with [O~I] $\lambda$5577, another O emission line which is detected in the optical spectra of SN\,2023ufx, following the method described in \cite{Taubenberger09}. After continuum subtraction, we rescale [O~I] $\lambda$5577 line to 1 / 1.14 of its initial intensity (i.e., the best-fit ratio of 6364 / 6300 from modeling [O~I] doublet), shift this scaled profile by 57 \AA\ (equivalent to the 64 \AA\ offset of the two lines in the [O~I] doublet) and add it to the original [O~I] $\lambda$5577 line profile. We find that such a modified [O~I] $\lambda$5577 can qualitatively reproduce the observed triple-peak line profile of [O~I] $\lambda \lambda$ 6300, 6364 in the nebular spectrum at day 211 (Figure \ref{fig:OIDoublet and OISinglet}). This clearly suggests that the observed three peaks in the [O~I] doublet is mostly comprised of O and not due to H$\alpha$ or other contamination. This supports a physical picture where two blobs of O are preferentially ejected along our line-of-sight.

Our 1-D line diagnostic tests are meant to be qualitative in nature to give an idea of the geometry of the explosion. A more quantitative analysis can be performed through 3-D simulations of various explosion models to re-create the observed line profiles like described in \cite{vanBaal24}, which is beyond the scope of this work. However, a consistent trend of blue and redshifted line profiles in SN\,2023ufx at nebular phases suggests an asymmetric distribution of ejecta from the explosion of a massive progenitor, possibly interacting with an asymmetrically distributed CSM. 
    
\section{Summary and Conclusions} \label{sec:8}

In this work, we report on our photometric and spectroscopic follow-up campaign of a luminous SN II, SN\,2023ufx. We obtained photometry from immediately after the explosion to $\sim$250 days after it. Our optical spectral coverage extended between days 7 and 211, while we obtained three NIR spectra between days 51 and 85. Here we summarize our results:

\begin{enumerate}
    \item SN\,2023ufx is a luminous ($M_{V}$ = -18.42 $\pm$ 0.08 mag) SNe\,II with the shortest observed plateau duration ($t_\mathrm{PT}$ $\sim$47 days) among SNe\,IIP. After a rise time to peak luminosity $\lesssim$7 days, a rapid decline is observed in all bands before the start of the plateau phase. This decline rate ($s1$ = 3.47 $\pm$ 0.09 mag / 50 days) is among the fastest across all SNe\,IIP and explores a new region in the peak luminosity vs. early decline phase space.
    \item By studying the radioactive decay tail of SN\,2023ufx, we estimate a total $^{56}$Ni mass of $\sim$0.14 $\pm$ 0.02 M$_{\odot}$. This is higher than $^{56}$Ni production in most SNe\,II. As the radioactive tail of SN\,2023ufx declines faster than the typical $^{56}$Co decay, we accounted for an incomplete trapping ($T_\mathrm{0}$ = 154.93 $^{+10.67}_{-10.89}$ days) of $\gamma$-rays. Some of the late-time luminosity might come from CSM interaction, in which case the estimated $^{56}$Ni could be overestimated.
    \item  The early optical spectra of SN\,2023ufx do not show flash features, suggesting that CSM around the progenitor if present is close by or asymmetrically distributed. Optical spectra of SN\,2023ufx at plateau phase differ from other SPSNe in two major ways: 1) While other SPSNe develop Fe~II absorption features early on in this phase, these features are extremely weak in SN\,2023ufx, even by the end of the plateau phase suggesting a low-metallicity ambient environment and 2) Even on the plateau, the strong and broad H$\alpha$ evolves showing evidence of multiple components. The low absorption-to-emission ratio of H$\alpha$ also suggests the presence of a small hydrogen envelope in SN\,2023ufx.
    \item Comparing the H emission features of the NIR Pa$\gamma$ and Pa$\beta$ with the asymmetric profile of H$\alpha$ confirms that the broad H$\alpha$ component is mostly due to hydrogen with little contamination from other elemental emission. 
    \item We compare the pseudo-bolometric light curve of SN\,2023ufx with a MESA+STELLA model grid with varying hydrogen envelope mass as one of the free parameters as described in \cite{Hiramatsu21}. We find that the extremely short plateau phase in SN\,2023ufx can be explained with a small hydrogen envelope of M$_\mathrm{H_\mathrm{env}}$ $\simeq$1.2 M$_{\odot}$. The best-fit model suggests a total mass at core-collapse to 8 -- 9.5 M$_{\odot}$ and thus a He-core mass of 6.8 -- 8.3 M$_{\odot}$, constraining the ZAMS mass of the progenitor of SN\,2023ufx to be M$_\mathrm{ZAMS}$ $\simeq$19 -- 25 M$_{\odot}$. We find that the early light curve ($\lesssim$10 days) cannot be explained with any of the models in the grid without introducing a total CSM mass of $\sim$0.09 M$_{\odot}$, suggesting elevated pre-explosion mass loss in the decades leading up to the SN explosion. A corresponding lack of flash features in the early spectra could be due the densest parts of the CSM being close enough to the progenitor where they get overrun by stellar ejecta soon after explosion and / or an asymmetrical distribution of CSM.
    \item We compared the two nebular spectra of SN\,2023ufx at days $\sim$188 and $\sim$211 days with an NLTE synthetic spectrum built by adding a hydrogen envelope to the Type Ib models discussed in \cite{Barmentloo24}. We find that a He-core mass of at least 6 M$_{\odot}$ is required in the synthetic spectra to match the nebular [O~I]~$\lambda\lambda$6300,\,6364 strength. This is in support of a high-mass progenitor scenario interpreted independently from comparisons between the psuedobolometric light curve and hydrodynamical model grids. Interestingly, none of the synthetic models can reproduce the observed strength of H$\alpha$ at late times suggesting that there could be possible ejecta CSM interactions. The flat-topped nature of the H$\alpha$ profile could indicate the presence of an asymmetric CSM distribution. 
    \item Strong emission lines [O~I]~$\lambda\lambda$6300,\,6364, H$\alpha$, and [Ca II]~$\lambda \lambda$7291,\,7323 evolve rapidly, forming multiple peaks by the nebular phase. Generally, all strong emission lines can be explained with a combination of ejecta moving preferentially along our line-of-sight. Part of the luminosity powering these emission lines could be due to ejecta interactions with an asymmetric distribution of CSM. Comparisons between the spatial distribution of [O~I]~$\lambda\lambda$6300,\,6364 and [O~I]~$\lambda$5577 shows that the observed triple peaks in the nebular spectra of SN\,2023ufx are primarily composed of O-ejecta. The flux ratio of [Ca~II] / [O~I] = 0.93 at late-times in SN\,2023ufx suggests a large O-core (i.e., a large progenitor mass) and that it may belong to an intermediate class between SNe\,II and SESNe. 
\end{enumerate}

In conclusion, SN\,2023ufx is a unique SN IIP with the shortest plateau length duration, potentially the result of an asymmetric explosion of a partially-stripped massive RSG likely interacting with an asymmetric distribution of CSM around it. The growing sample of SPSNe like SN\,2023ufx will be crucial to understand the evolution of massive RSG progenitors of SNe\,II, by constraining late-stage mass-loss in core-collapse SNe, understanding their progenitor channels, and explosion physics.
    
\section*{Acknowledgments}
Research by A.P.R., S.V., Y.D., N.M.R, E.H., and D.M. is supported by NSF grant AST-2407565. 
Time-domain research by the University of Arizona team and D.J.S. is supported by National Science Foundation (NSF) grants 2108032, 2308181, 2407566, and 2432036 and the Heising-Simons Foundation under grant \#2020-1864.  
This work makes use of data from the Las Cumbres Observatory global telescope network, which is supported by NSF grants AST-1911225 and AST-1911151.

This publication was made possible through the support of an LSST-DA Catalyst Fellowship to K.A.B., funded through Grant 62192 from the John Templeton Foundation to LSST Discovery Alliance. The opinions expressed in this publication are those of the authors and do not necessarily reflect the views of LSST-DA or the John Templeton Foundation.
Supernova research at Rutgers University is supported in part by NSF award 2407567 to S.W.J.

R.K.T. is supported by the NKFIH/OTKA FK-134432 and the NKFIH/OTKA K-142534 grant of the National Research, Development and Innovation (NRDI) Office of Hungary.
A.V.F.'s research group at UC Berkeley acknowledges financial assistance from the Christopher R. Redlich Fund, as well as donations from Gary and Cynthia Bengier, Clark and Sharon Winslow, Alan Eustace, William Draper, Timothy and Melissa Draper, Briggs and Kathleen Wood, and Sanford Robertson (W.Z. is a Bengier-Winslow-Eustace Specialist in 
Astronomy, T.G.B. is a Draper-Wood-Robertson Specialist in Astronomy, Y.Y. was a Bengier-Winslow-Robertson Fellow in Astronomy) and numerous other donors.

CPG acknowledges financial support from the Secretary of Universities and Research (Government of Catalonia) and by the Horizon 2020 Research and Innovation Programme of the European Union under the Marie Sk\l{}odowska-Curie and the Beatriu de Pin\'os 2021 BP 00168 programme, from the Spanish Ministerio de Ciencia e Innovaci\'on (MCIN) and the
Agencia Estatal de Investigaci\'on (AEI) 10.13039/501100011033 under the
PID2023-151307NB-I00 SNNEXT project, from Centro Superior de
Investigaciones Cient\'ificas (CSIC) under the PIE project 20215AT016
and the program Unidad de Excelencia Mar\'ia de Maeztu CEX2020-001058-M,
and from the Departament de Recerca i Universitats de la Generalitat de
Catalunya through the 2021-SGR-01270 grant.

Some of the data presented herein were obtained at Keck Observatory, which is a private 501(c)3 non-profit organization operated as a scientific partnership among the California Institute of Technology, the University of California, and the National Aeronautics and Space Administration. The Observatory was made possible by the generous financial support of the W. M. Keck Foundation. The authors wish to recognize and acknowledge the very significant cultural role and reverence that the summit of Maunakea has always had within the indigenous Hawaiian community.  We are most fortunate to have the opportunity to conduct observations from this mountain.

The LBT is an international collaboration among institutions in the United States, Italy and Germany. LBT Corporation Members are: The University of Arizona on behalf of the Arizona Board of Regents; Istituto Nazionale di Astrofisica, Italy; LBT Beteiligungsgesellschaft, Germany, representing the Max-Planck Society, The Leibniz Institute for Astrophysics Potsdam, and Heidelberg University; The Ohio State University, and The Research Corporation, on behalf of The University of Notre Dame, University of Minnesota and University of Virginia.

A major upgrade of the Kast spectrograph on the Shane 3\,m telescope at Lick Observatory, led by Brad Holden, was made possible through generous gifts from the Heising-Simons Foundation, William and Marina Kast, and the University of California Observatories. Research at Lick Observatory is partially supported by a generous gift from Google. 

This paper made use of the modsCCDRed data reduction code developed in part with funds provided by NSF Grants 
AST-9987045 and AST-1108693. 
%
This research made use of the NASA/IPAC Extragalactic Database (NED), which is funded by the National Aeronautics and Space Administration and operated by the California Institute of Technology.

\facilities{Las Cumbres Observatory, W. M. Keck, Lick Observatory, MMT, Large Binocular Telescope (LBT), Hobby-Eberly Telescope, Neil Gehrels Swift Observatory,  Bok (B\&C), Southern Astrophysical Research Telescope (SOAR), ATLAS, and ZTF}

\software{Astropy \citep{Astropy13, Astropy18}, Matplotlib \citep{Hunter07}, Numpy \citep{Harris20}, Scipy \citep{Virtanen20}, IRAF \citep{Tody86}, \texttt{xtellcorr} \citep{Vacca03}, \texttt{lcogtsnpipe} \citep{Valenti16}} 

\appendix

\section{Spectroscopy Observation Log}

\startlongtable
\begin{deluxetable*}{cccccc}
\tablenum{A1}
\tablecaption{Optical and NIR Spectra of SN\,2023ufx\label{tab:optical_nir_spectralog}}
\tablewidth{0pt}
\tablehead{
\colhead{UTC Date \& Time (hh:mm)} & \colhead{Modified Julian Date (Days)} & \colhead{Phase (Days)} &
\colhead{Telescope} & \colhead{Instrument} & \colhead{Wavelength Coverage (\AA)} }
\startdata 
 2023-10-13 13:15 & 60230.55 &  7& FTN & FLOYDS &  3500 -- 10000\\
 2023-10-13 14:02 & 60230.59 & 7 & Keck I & LRIS & 3148 -- 10086\\ 
 2023-10-13 14:52 & 60230.62 & 7 & Keck I & LRIS & 3147 -- 10086\\
 2023-10-17 14:38 & 60234.61 &  11 & FTN & FLOYDS & 3500 -- 10000\\
 2023-10-19 10:30 & 60236.44 &  13 & MMT & Binospec & 3826 -- 9197\\
 2023-10-19 13:59 & 60236.58 &  14 & FTN & FLOYDS & 3500 -- 10000\\
 2023-10-20 14:09 & 60237.59 &  15 & FTN & FLOYDS & 3500 -- 10000\\
 2023-10-24 13:02 & 60241.54 &  18 & FTN & FLOYDS & 3500 -- 9999\\
 2023-10-28 12:45 & 60245.53 &  22 & FTN & FLOYDS & 3500 -- 10000\\
 2023-11-01 12:59 & 60249.54 &  26 & FTN & FLOYDS & 3500 -- 9999\\
 2023-11-08 12:04 & 60256.50 &  33 & FTN & FLOYDS & 3500 -- 9999\\
 2023-11-09 10:21 & 60257.43 & 34 & LBT & MODS & 3500 -- 10000\\
 2023-11-10 04:45 & 60258.20 & 35 & Shane & Kast & 3636 -- 10734\\
 2023-11-12 11:48 & 60260.49 & 37 & FTN & FLOYDS & 3500 -- 10000\\
 2023-11-13 15:28 & 60261.61 &  38 & Keck I & LRIS & 3139 -- 10216 \\
 2023-11-16 11:32 & 60264.48 & 41 & FTN & FLOYDS & 3302 -- 10000\\
 2023-11-20 12:15 & 60268.51 & 45 & FTN& FLOYDS & 3500 -- 9999\\
 2023-11-26 10:56 & 60274.46 & 51 & FTN& FLOYDS & 3500 -- 9999\\
 2023-12-03 10:27 &  60281.43 &  58 & MMT & MMIRS & 9927 -- 15015\\
 2023-12-04 13:19 & 60282.55 & 59 & FTN& FLOYDS & 3500 -- 10000\\
 2023-12-12 04:13 & 60290.43 & 67 & Shane & Kast & 3640 -- 8744 \\
 2023-12-12 11:35 & 60290.48 & 67 & Keck I & LRIS & 3139 -- 10064\\
 2023-12-18 06:32 &  60296.27 &  73 & SOAR & TripleSpec & 9429 -- 17913 \\
 2023-12-18 13:44 & 60296.57 & 73 & FTN & FLOYDS & 3499 -- 10000\\
 2023-12-26 15:07 & 60304.63 & 81& FTN & FLOYDS & 3500 -- 9999\\
 2023-12-30 13:23 &  60308.55 &  85 & Keck II & NIRES & 9408 -- 24693\\
 2023-12-31 10:06 & 60309.42& 86& FTN & FLOYDS & 3500 -- 10000\\
 2024-01-06 12:40 & 60315.53 & 93 & FTN & FLOYDS & 3500 -- 10000\\
 2024-01-22 12:03 & 60331.50& 109&FTN & FLOYDS & 3500 -- 9999\\
 2024-01-28 05:04 & 60337.21 & 114 & SOAR & Goodman-RED & 4934 -- 8938\\
 2024-02-03 10:08 & 60343.42 & 120&FTN & FLOYDS & 3500 -- 10000\\
 2024-02-07 10:56 & 60347.46 & 124 & Keck I & LRIS & 3139 -- 10258\\
 2024-02-09 12:10 & 60349.51 & 127&FTN & FLOYDS & 3500 -- 10000\\
 2024-02-13 04:03 & 60353.2 & 130 & Bok & B\&C & 4000 -- 7998\\
 2024-02-19 09:16 & 60359.39 & 136&FTN & FLOYDS & 3499 -- 10000\\
 2024-02-27 08:51 & 60367.37 & 144&FTN & FLOYDS & 3500 -- 9999\\
 2024-03-17 03:17 & 60386.14 & 163 & Shane & Kast & 3624 -- 10748\\
 2024-03-18 09:16 & 60387.39 & 164&FTN & FLOYDS & 3499 -- 9000\\
 2024-04-11 02:22 & 60411.10 & 188 & Shane & Kast & 3624 -- 10756\\
 2024-05-04 02:45 & 60434.11 & 211 & HET & LRS2-B & 3640 -- 6945
\enddata{}
\end{deluxetable*}

\end{document}